\documentclass[letterpaper,usenatbib]{mnras}
\usepackage{graphicx}
\usepackage{color}
\usepackage{lscape}
\usepackage[figuresright]{rotating}
\usepackage{longtable}
\usepackage{mathrsfs,amsmath}
\usepackage{epstopdf}
\newsavebox{\tablebox}
\usepackage{hyperref}
\usepackage{times}
\usepackage{longtable,lscape}

\newcommand{\lv}{\ifmmode \lambda L_{\lambda}(5100~\rm \AA) \else $\lambda L_{\lambda}(5100\AA)$\ \fi}
\newcommand{\kms}{\ifmmode {\rm km\ s}^{-1} \else km s$^{-1}$\ \fi}
\newcommand{\ergs}{\ifmmode {\rm erg\ s}^{-1} \else erg s$^{-1}$\ \fi}
\newcommand{\lb}{\ifmmode L_{\rm Bol} \else $L_{\rm Bol}$\ \fi}
\newcommand{\ledd}{\ifmmode L_{\rm Edd} \else $L_{\rm Edd}$\ \fi}
\newcommand{\hb}{\ifmmode H\beta \else H$\beta$\ \fi}
\newcommand{\ha}{\ifmmode H\alpha \else H$\alpha$\ \fi}
\newcommand{\sil}{\ifmmode \sigma_{\rm line} \else $\sigma_{\rm line}$\ \fi}
\newcommand{\mbh}{\ifmmode M_{\rm BH}  \else $M_{\rm BH}$\ \fi}
\newcommand{\msun}{M_{\odot}}
\newcommand{\rfe}{\ifmmode R_{\rm Fe} \else $R_{\rm Fe}$\ \fi}
\newcommand{\sst}{\ifmmode \sigma_{\rm \ast}\else $\sigma_{\rm \ast}$\ \fi}
\newcommand{\dhb}{\ifmmode D_{\rm H\beta} \else $D_{\rm H\beta}$\ \fi}
\newcommand{\leddR}{\ifmmode L_{\rm Bol}/L_{\rm Edd} \else $L_{\rm Bol}/L_{\rm Edd}$\ \fi}
\newcommand{\ms}{\ifmmode M_{\rm BH}-\sigma_{\ast} \else $M_{\rm BH}-\sigma_{\ast}$\ \fi}
\newcommand{\sm}{\ifmmode \sigma_{\rm H\beta,mean} \else $\sigma_{\rm H\beta,mean}$\ \fi}
\newcommand{\sr}{\ifmmode \sigma_{\rm H\beta,rms} \else $\sigma_{\rm H\beta,rms}$\ \fi}
\newcommand{\fwm}{\ifmmode \rm FWHM_{\rm mean} \else $\rm FWHM_{\rm mean}$\ \fi}
\newcommand{\fwr}{\ifmmode \rm FWHM_{\rm rms} \else $\rm FWHM_{\rm rms}$\ \fi}
\newcommand{\vpfm}{\ifmmode \rm VP_{\rm F, mean} \else $\rm VP_{\rm F, mean}$\ \fi}
\newcommand{\vpsm}{\ifmmode \rm VP_{\rm \sigma, mean} \else $\rm VP_{\rm \sigma, mean}$\ \fi}
\newcommand{\vpfr}{\ifmmode \rm VP_{\rm F, rms} \else $\rm VP_{\rm F, rms}$\ \fi}
\newcommand{\vpsr}{\ifmmode \rm VP_{\rm \sigma, rms} \else $\rm VP_{\rm \sigma, rms}$\ \fi}
\newcommand{\feii}{Fe {\sc ii}\ }
\begin{document}
\title[Calibration of the factor $f$ in SMBH masses]{Calibration of the virial factor $f$ in supermassive black hole masses of reverberation-mapped AGNs }
\author[L. Yu et al.]{Li-Ming Yu$^{1}$, Wei-Hao Bian$^{1}$ \thanks{E-mail: whbian@njnu.edu.cn}, Chan Wang$^{1}$, Bi-Xuan Zhao$^{1}$, Xue Ge$^{1}$ \\
$^1$School of Physics and Technology, Nanjing
Normal University, Nanjing 210046, China\\
}\maketitle

\begin{abstract}

Using a compiled sample of 34 broad-line active galactic nuclei (AGNs) with measured \hb time lags from the reverberation mapping (RM) method and measured bulge stellar velocity dispersions \sst, we calculate the virial factor $f$ by assuming that the RM AGNs intrinsically obey the same \ms relation as quiescent galaxies, where \mbh is the mass of the supermassive black hole (SMBH). Considering four tracers of the velocity of the broad-line regions (BLRs), i.e., the \hb line width or line dispersion from the mean or rms spectrum, there are four kinds of the factor $f$. Using the \hb Full-width at half-maximum (FWHM) to trace the BLRs velocity, we find significant correlations between the factor $f$ and some observational parameters, e.g., FWHM, the line dispersion. Using the line dispersion to trace the BLRs velocity, these relations disappear or become weaker. It implies the effect of inclination in BLRs geometry. It also suggests that the variable $f$ in \mbh estimated from luminosity and FWHM in a single-epoch spectrum is not negligible. Using a simple model of thick-disk BLRs, we also find that, as the tracer of the BLRs velocity, \hb FWHM has some dependence on the inclination, while the line dispersion $\sigma_{\rm \hb}$ is insensitive to the inclination. Considering the calibrated FWHM-based factor $f$ from the mean spectrum, the scatter of the SMBH mass is 0.39 dex for our sample of 34 low redshift RM AGNs. For a high redshift sample of 30 SDSS RM AGNs with measured stellar velocity dispersions, we find that the SMBH mass scatter is larger than that for our sample of 34 low redshift RM AGNs. It implies the possibility of evolution of the \ms relation from high-redshift to low-redshift AGNs.


\end{abstract}


\section{INTRODUCTION}
It is believed that active galactic nuclei (AGNs) are powered by accretion mass onto the central supermassive black holes (SMBHs), which are also assumed to exist in the center of all quiescent galaxies \citep{KR95}. There are mainly two parameters for SMBHs, i.e., mass (\mbh) and spin, which need to be determined. For a few very nearby ($<$ 100 Mpc) quiescent galaxies, including our Galaxy, SMBHs masses can be measured through the stellar or gaseous dynamics method \citep[e.g.][]{Tr02,Mc11}. It has been found the nearby quiescent galaxies follow a tight correlation between the central SMBHs mass and the bulge or spheroid stellar velocity dispersion (\sst) called \ms relation
 \citep{FM00,Ge00,KH13},
 \begin{equation}
 \label{eq1}
\log \frac{\mbh (\sigma_*)}{10^9 \msun}=\alpha+\beta \log \frac{\sst}{~200\kms},
\end{equation}
Considering different definition of \sst and different bulge type, there are different values of index $\beta$ and intercept $\alpha$ \citep[e.g.][]{KH13}.

Because of a limit of space resolution and  outshining the host for faraway AGNs, it is much more difficult to derive the SMBHs masses through stellar or gaseous dynamics method. AGNs can be classified into type 1 or type 2 AGNs, depending on whether the broad-line regions (BLRs) can be viewed directly. For type 1 AGN, BLR can be used as a probe of the gravitational potential of the SMBHs, and SMBHs masses can be derived as follows,
\begin{equation}
\label{eq2}
\mbh=f\frac{R_{\rm BLR}~(\Delta V)^2}{G} \equiv f \rm{VP},
\end{equation}
where $R_{\rm BLR}$ is the distance from black hole to the BLRs, $\Delta V$ is the velocity of BLRs clouds, G is the gravitational constant. $\Delta V$ is usually traced by the Full-width at half-maximum (FWHM) or the line dispersion ($\sigma_{\rm \hb}$) of \hb emission line measured from the mean or rms spectrum. $\rm VP$ is the so-called virial product,  $\rm VP=R_{\rm BLR}~(\Delta V)^2/{G}$. $f$ is a virial coefficient to characterize the kinematics, geometry, inclination of the BLRs clouds. For type 2 AGNs, torus obscures BLRs in the line of sight, which makes broad lines invisible, except in some of the polarization spectra \citep[e.g.][]{Zhang2008}. Using the \ms relation, the SMBH masses in type II AGNs are mainly derived from \sst \citep{Ka03a,Ka03b,Bian2006}.

Considering the photon-ionization model of BLRs in AGNs, $R_{\rm BLRs}$ can be estimated from the reverberation mapping (RM) method \citep[e.g.][]{BM82,Pe93}. The variation in the continuum would lead to the changes in the broad-line emission delayed by a lag time $\tau=R_{\rm BLRs}/c$. The lag time of $\tau$ was successfully measured by the RM method for nearly 120 AGNs \citep[e.g.][]{Pe04, Gr17b,Du18}. For RM AGNs, an empirical $R_{\rm BLR}-L_{\rm 5100}$ relation is also suggested, where $L_{\rm 5100}$ is the continuum luminosity at 5100 \AA\ \citep{Ka00, Ka05, Be13,Du18}. The complexity of this relation is beyond the scope of this paper \citep[e.g.][]{Du18}.

Comparing the scatter of different VP based on the line dispersion or FWHM of \hb measured from the rms or mean spectrum, \cite{Pe04} preferred the VP based on $\sigma_{\rm \hb}$ from the rms spectrum. $f$ was therefore assumed as a constant. However, $f$ may be different for each object. Assuming BLRs velocity distribution is isotropic, $f=0.75$ when $\Delta V$ is traced by \hb FWHM \citep{Ne90}.  Given the complicated structures of BLRs inferred from the velocity-binned RMs,  $f$ is most likely to vary from object to object \citep[e.g.][]{Xiao2018}. By modelling simultaneously the AGNs continuum light curve and \hb line profiles, some BLRs dynamical models  found that there was a wide range of $f$ and $f$ has a correlation with the inclination angle, or \mbh\ \citep[e.g.][]{Pancoast2014,Gr17a, Williams2018, Pancoast2018, Li2018}. It was suggested that the geometry of BLRs was thick disk, the kinematics of BLRs can de described by elliptical orbits or inflowing or outflowing trajectory \citep{Williams2018}. $f$ can also be derived through the \mbh calculated from resolved Pa$\alpha$ emission region by Very Large Telescope Interferometry \citep[VLTI,][]{Sturm2018}, the accretion disk model fit of AGNs spectral energy distribution (SED) \citep{Mejia2018}, or X-ray variability\citep{Pan2018}, which are assumed to be independent of inclination. For the SMBHs binary with their own BLRs, complexity of the broad-line profiles was suggested, which would lead to the complexity of $f$ \citep[e.g.][]{SS2019}.

If AGNs follow the same \ms relation as quiescent galaxies, scaling the RM AGNs with measured stellar velocity dispersion \sst to the quiescent galaxies yields an empirical calibration of the average value of the factor $f$ in Equation \ref{eq1}. This is a commonly approach to derive the mean value $f$. Based on the early \ms relation \citep{FM00,Tr02}, \cite{On04} suggested $f_{\rm \sigma,rms}$=5.5 for $\sigma_{\rm \hb}$-based VP from the rms spectrum for a sample of 16 RM AGNs. With larger samples, others also found a consistent $f$ \citep[e.g.][]{Wo10, Gr11, Gr13, HK14}. In addition to the average $f$, it was also suggested a systematically smaller $f$ for AGNs with bars/pseudobulges \citep[e.g.][]{Gr11, HK14}.

Based on a sample of 14 RM AGNs with measured \sst, \cite{Co06} suggested that $f$ based on FWHM($\rm \hb$) has a dependence on FWHM($\rm \hb$), $\sigma_{\rm \hb}$, and their ratio $D_{\rm \hb}=\rm FWHM(\hb)/\sigma_{\hb}$. For a sample of 24 RM AGNs with \sst including Lick AGN RM project, \cite{Wo10} found that there were no strong correlations between $f$ and parameters, such as the Eddington ratio or line-shape measurements (see their Figure 6).




In this paper, using a larger sample of 34 RM AGNs with measured \sst, we calculate the factor $f$ from RM VP and $\sst$-based \mbh. We investigate the relation between the factor $f$ and other observational parameters. Section 2 presents our sample. Section 3 is data analysis and Section 4 is discussion. Section 5 summaries our results. All of the cosmological calculations in this paper assume
$\Omega_{\Lambda}=0.7$, $\Omega_{\rm M}=0.3$, and $H_{0}=70~ \kms {\rm Mpc}^{-1}$.


\section{SAMPLE}

Up to now, there are about 120 AGNs with measured \hb/\ha lags from the RM method \citep[e.g.][]{Gr17b,Du18}. Our sample consists of 34 low redshift broad-line AGNs ($z$ less  than $0.1$ except PG 1617+175) with both measured \hb lags and reliable \sst, which allows us to calibrate the factor $f$ based on the \ms relation. 32 out of these 34 RM AGNs are selected from \cite{HK14}, who had imaged these objects and classified them into three bulge types: elliptical, classical bulges and pseudobulge. For Fairall 9, the stellar velocity dispersion is adopted from its near-infrared spectrum \citep{O95}. Beyond the sample of \cite{HK14}, there are two additional objects. The first one is an early-type galaxy NGC 5273 from \cite{Be14}. The second one is MCG+06-26-01 with pseudobulge. Its RM result is from \cite{Du15} and its stellar velocity dispersion is adopted from \cite{Wo15}.   For our sample of these 34 RM AGNs, there are 8 classified as ellipticals, 9 classified as classical bulges, 17 classified as pseudobulges.  Here we do not use a high-z sample of 44 AGNs ($z~\sim 0.1-1.0$) from the Sloan Digital Sky Survey (SDSS) RM Project \cite{Gr17b} with measured \hb/\ha lags to do the calibration of $f$. There are 30 out of 44 AGNs with measured \sst by \cite{Sh15}. We use this high-z sample to investigate the evolution of the \ms relation.


In this paper, considering different host type of RM AGNs, the relations between the factor $f$ and observational parameters are investigated, as well as the BLRs physics. For the \hb emission line, there are four kinds of parameters to trace the BLRs velocity $\Delta V$ from the mean and rms spectrum, i.e., \fwm, \sm, \fwr, \sr, respectively. The ratio of the FWHM($\rm \hb$) to  $\sigma_{\rm \hb}$ is defined as \dhb, i.e., \dhb(mean)=\fwm/\sm, \dhb(rms)=\fwr/\sr. The value of \dhb is 2.35, 3.46, 2.45, 2.83, and 0 for a Gaussian, a rectangular, a triangular, an edge-on rotating ring, and a Lorentzian profile, respectively. \dhb depends on the line profile and gives a simple, convenient parameter that may be related to the dynamics of the BLRs \citep{Co06, Du16}. The Eddington ratio \lb/\ledd is an important parameter describing the SMBH accretion process, where \lb and \ledd are the bolometric luminosity and the Eddington luminosity, respectively. It depends on the estimations of \mbh($\sigma_*$) and \lb, where $\lb=k_{5100} \times \lv$, and $k_{5100}=9$. The bolometric correction coefficient $k_{5100}$ was suggested to have a dependence on the luminosity or the Eddington ratio \citep[e.g.][]{Marconi2004, Jin2012}. \rfe (the ratio of optical \feii and \hb flux) has a correlation with Eigenvector 1 (EV1) from principal component analysis (PCA), which has been demonstrated to be driven by the Eddington ratio \citep{BG92, Su00, Sh14, Ge2016}. We can also use \rfe to demonstrate the Eddington ratio.

Properties about our sample of 34 RM AGNs for calibration are presented in Table \ref{tab1}. Col. (1-2) give the source name and the alternate name. Col. (3) is \rfe measured by \cite{Du16}. Col. (4) is the stellar velocity dispersion \sst. Col. (5) is the bulge type, E is elliptical, CB is classical bulge and PB is pseudobulge. Col. (6) is host-corrected monochromatic luminosity at 5100 \AA. Col. (7) is the rest-frame \hb lag in units of days. Col. (8) is the broad \hb \fwm from the mean spectrum. Col. (9) is the broad \hb \fwr measured from the rms spectrum. Col.(10) is the broad \hb line dispersion \sm from the mean spectrum.  Col. (11) is the broad \hb line dispersion \sr measured from the rms spectrum. Col. (12) is virial product calculated based on  \fwm. Col. (13) is virial product calculated based on \sm.  Col. (14) is virial product calculated based on \fwr. Col. (15) is virial product calculated based on \sr. Col. (16) is references for these data \citep[i.g.][]{O95, Co06, Wang2014, Be14,HK14, Wo15, Du15, Du16, Williams2018}.  Col.(4)-Col.(11) are mainly from \cite{HK14},  the remaining data come from \cite{Williams2018} marked with superscript a, \cite{Ba15} marked with superscript b,  table 1 in \cite{Du16}  marked with superscript c, table 7 in \cite{Du15} marked with superscript d, table 1 in \cite{Co06} marked with superscript e, \cite{Be14} marked with superscript f, \cite{Wo15} marked with superscript g, \cite{Wang2014} marked with superscript h, and \cite{O95} marked with superscript i. For the high-z sample of 30 SDSS RM AGNs, we also present their $z$, \sst, four kinds of velocity of \hb/\ha and the corresponding VP in Table  \ref{tab2}. There are 10 AGNs with both measurements of \ha and \hb lags.

\section{Data Analysis}

\subsection{Distributions of the factor $f$}
\begin{figure*}
\includegraphics[angle=-90,width=3.0in]{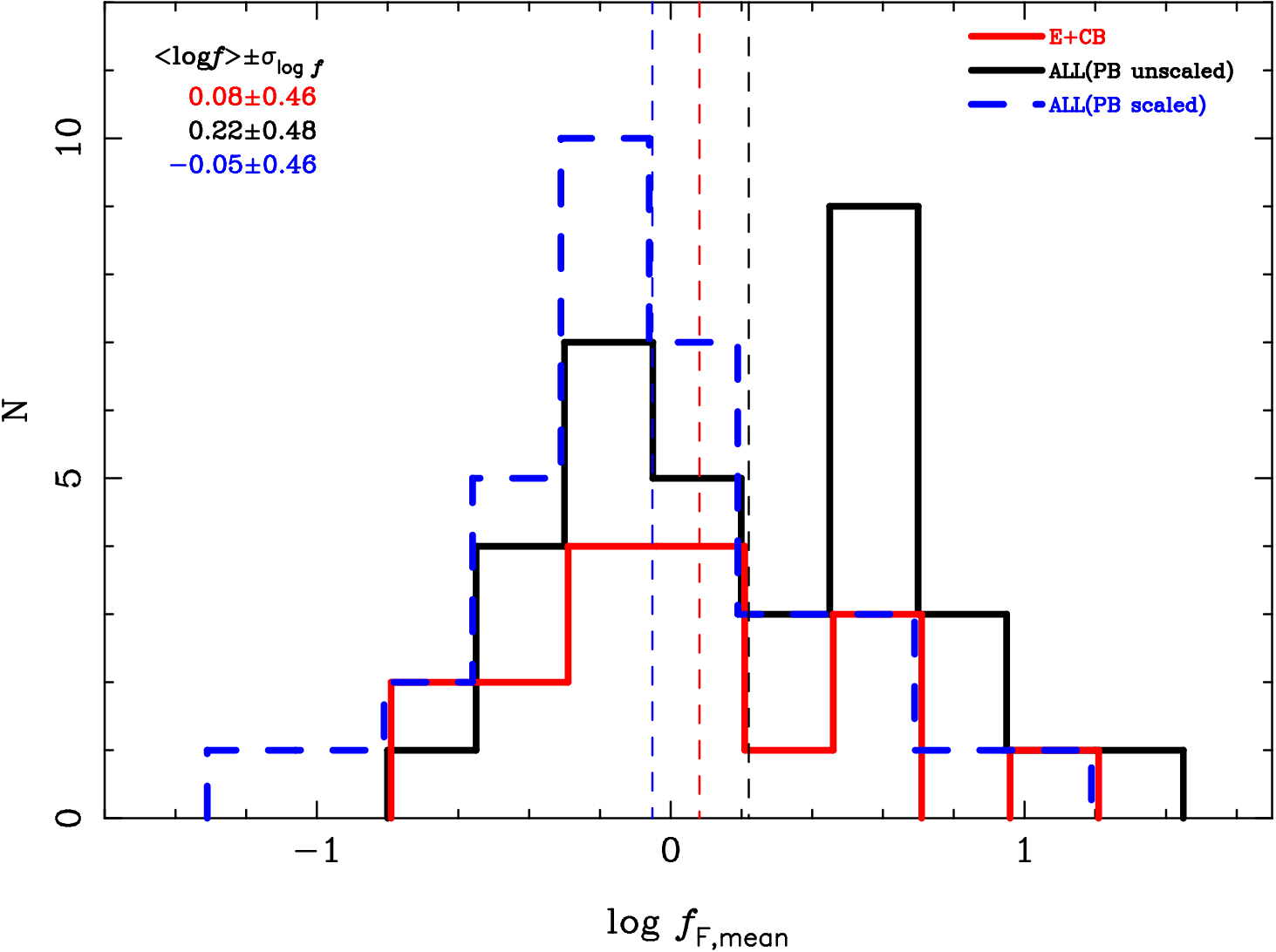}
\includegraphics[angle=-90,width=3.0in]{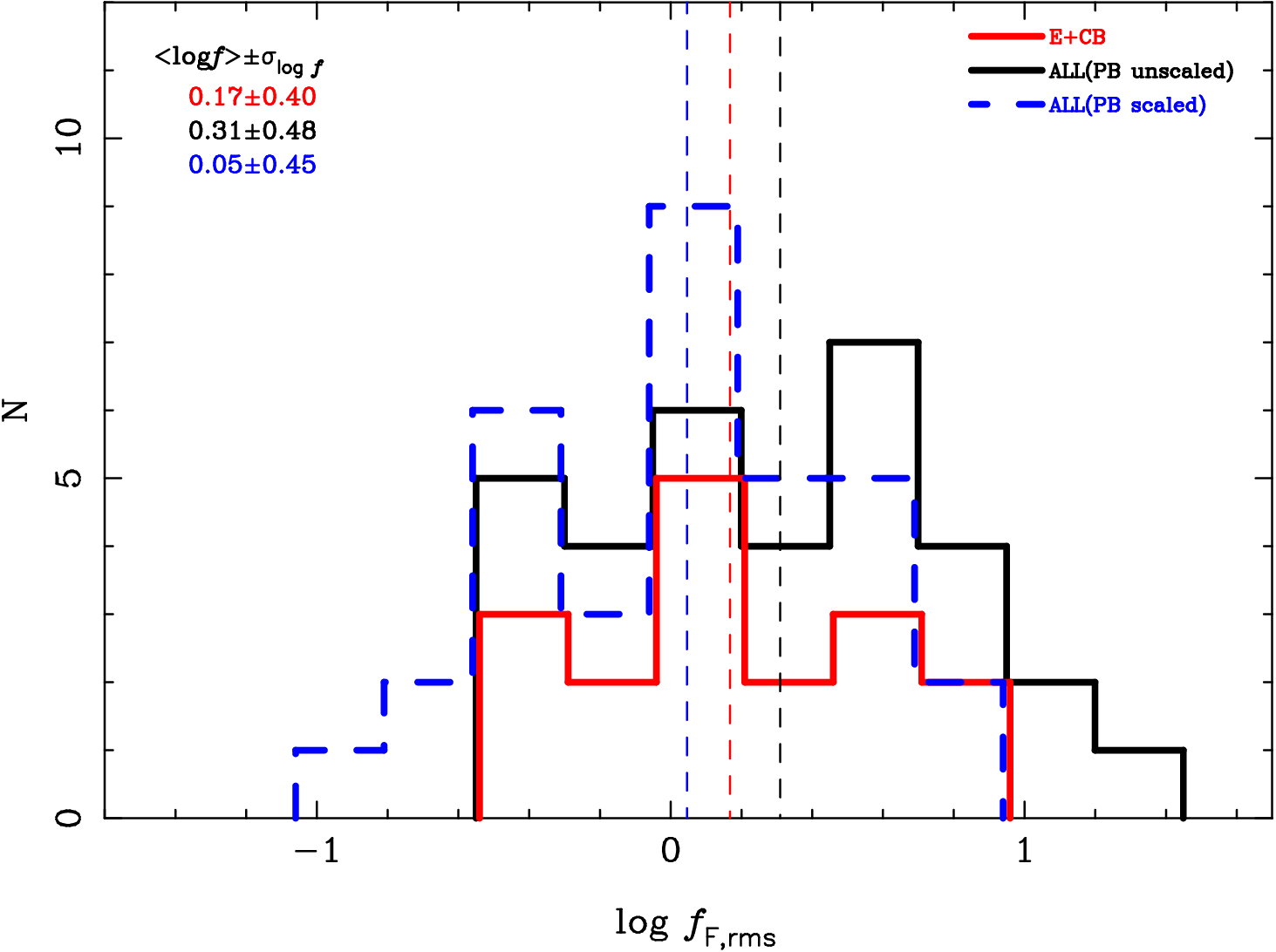}
\includegraphics[angle=-90,width=3.0in]{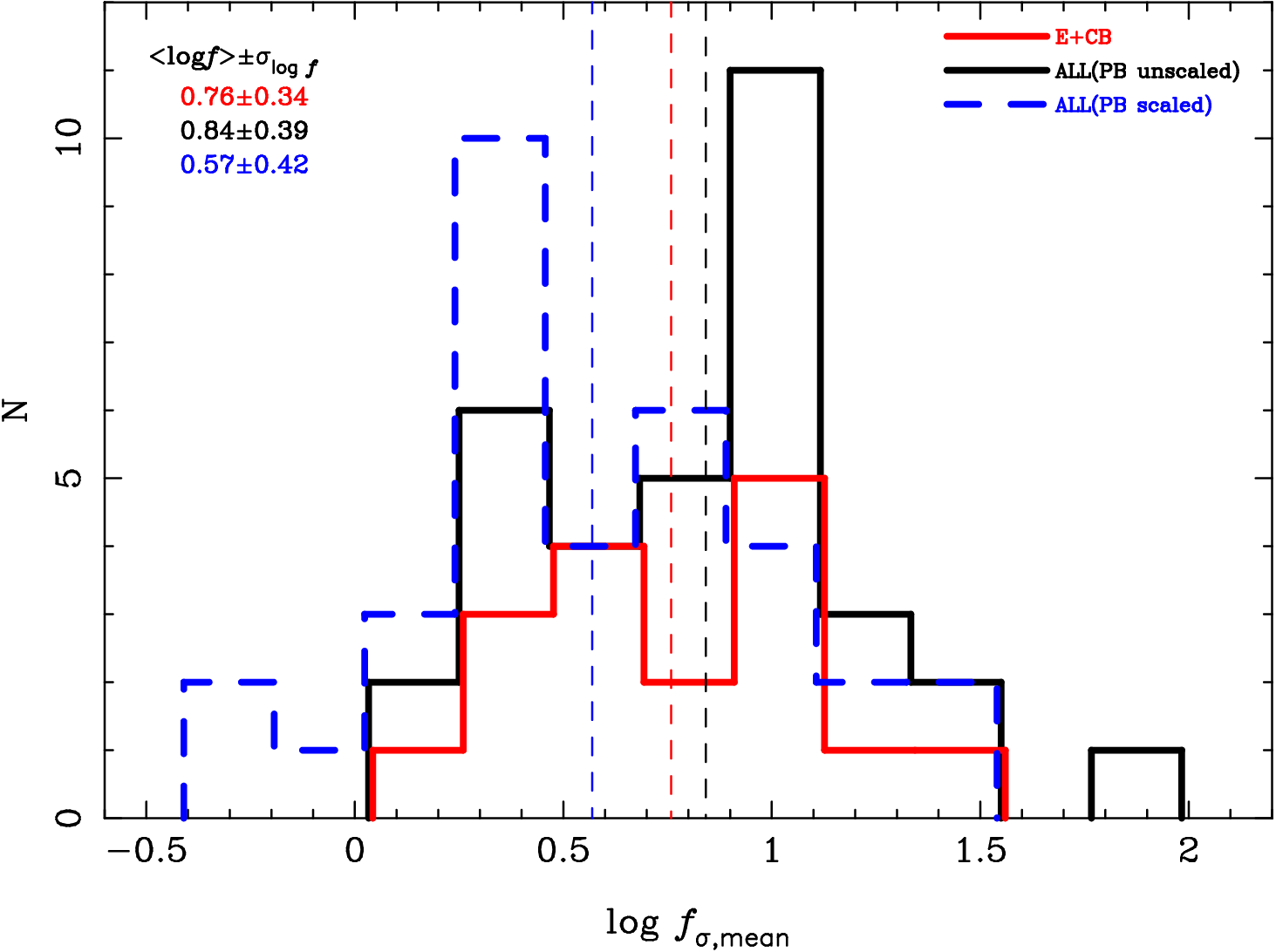}
\includegraphics[angle=-90,width=3.0in]{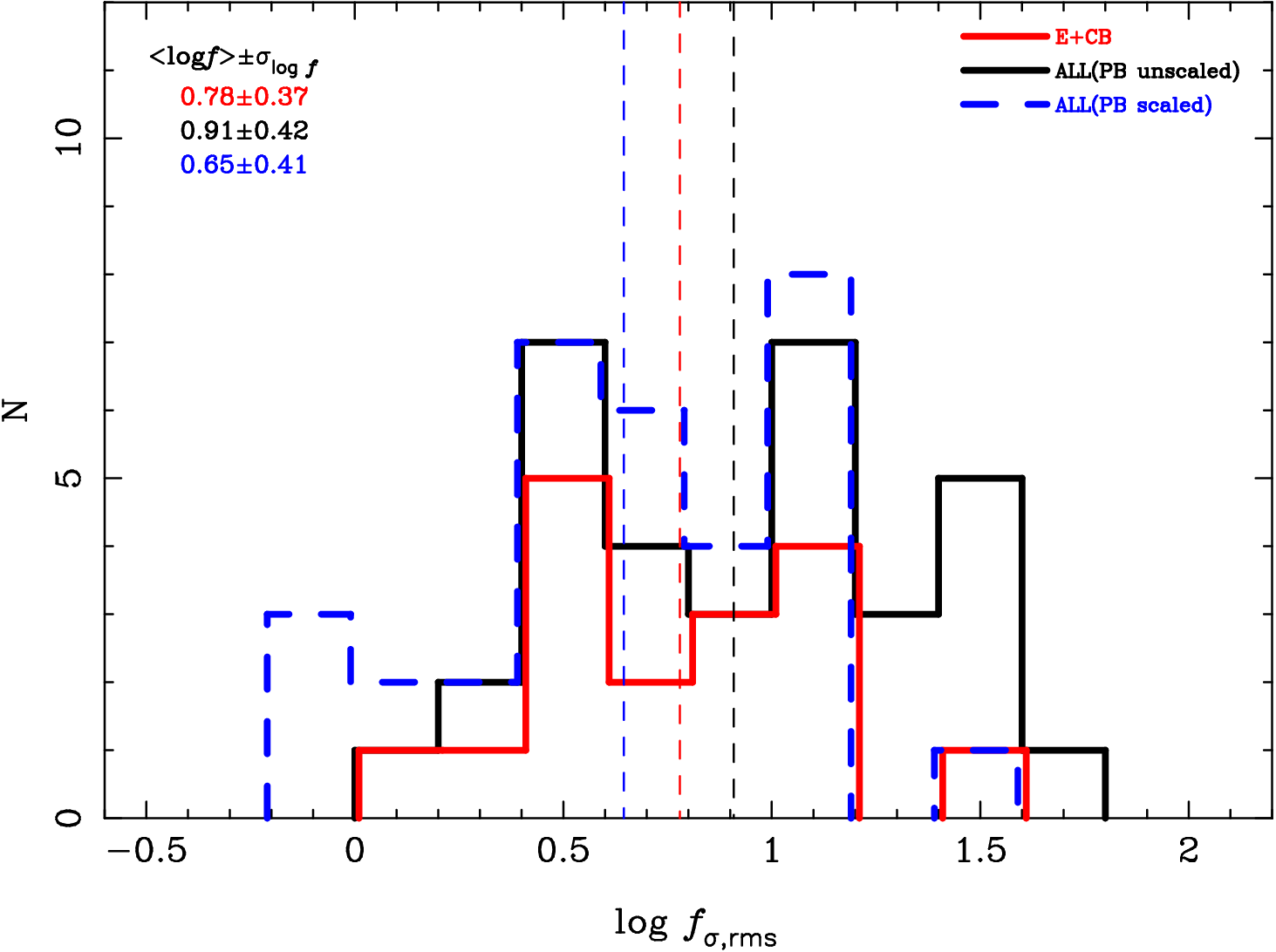}
\caption{The distributions of four kinds of the factor $f$, i.e., $f_{\rm F,mean}, f_{\rm F,rms}, f_{\rm \sigma,mean}, f_{\rm \sigma,rms}$ from left to right and from top to bottom, respectively. The red line is the $f$ distribution for our sample excluding pseudobulges. The black line and the  blue dash-line are the $f$ distributions for our total sample adopting different \ms relation for pseudobulges with $\alpha=-0.51$ (PB unscaled $f$) and $\alpha=-1.09$ (PB scaled $f$ by a factor of 3.80), respectively. The vertical dashed-line denote the mean value of $f$ for these three kinds of distributions. The mean value and the standard deviation of $f$ are shown in each panel.}
\label{Fig1}
\end{figure*}

To calibrate the virial factor $f$ in Equation \ref{eq2}, we need other independent SMBH mass measurements for these RM AGNs. Using the \ms relation of quiescent galaxies \citep{KH13}, we derive the black hole masses for our sample, except NGC3227 and NGC 4151. For NGC 3227 and NGC 4151, their SMBHs masses were give through the stellar dynamic method \citep{Da06, On14}. For coefficients in the \ms relation shown in Equation \ref{eq1}, $\beta=4.38\pm 0.29, \alpha=-0.51\pm 0.05$ for ellipticals and classical bulges \citep{KH13}, and $\beta=4.38\pm 0.29, \alpha=-1.09\pm 0.05$ for pseudobulge  \citep{HK14}. For pseudobulge with the same \sst, the \sst-based \mbh is smaller and scaled by a factor of $10^{[1.09-0.51]}=3.80$ when using $\alpha=-1.09$ instead of $\alpha=-0.51$. With available \sst, we calculate SMBHs masses, and then calculate the virial factor $f$ for individual source from Equation \ref{eq2}. There are four kinds of $\rm VP$ from four kinds of tracers of $\Delta V$, i.e., \vpfm, \vpsm, \vpfr, \vpsr calculated from \hb \fwm, \sm, \fwr, \sr, respectively. Therefore, there are four kinds of $f$ as follows,
\begin{equation}\label{eq3}
\begin{split}
f_{\rm F,mean} & =\frac{\mbh(\sigma_*)}{\vpfm},
f_{\rm \sigma, mean}=\frac{\mbh(\sigma_*)}{\vpsm},\\
f_{\rm F, rms} & =\frac{\mbh(\sigma_*)}{\vpfr},
f_{\rm \sigma, rms}=\frac{\mbh(\sigma_*)}{\vpsr},
\end{split}
\end{equation}
For pseudobulge with the same \sst, the factor $f$ is smaller and scaled by a factor of 3.80 when using smaller $\alpha$ and same slope of $\beta=4.38$ in the \ms relation, i.e., $\alpha=-0.51$ (PB unscaled $f$), and $\alpha=-1.09$ (PB scaled $f$ by a factor of 3.80) \citep{HK14}. The errors of virial coefficient $\log f$ can be calculated from the errors of $\rm VP$ and \mbh, i.e., the errors of the $\Delta V$, the \hb lag and \sst as follows (only including the observational uncertainties),
\begin{equation}\label{eq4}
\delta \log f=\sqrt{(\delta \log VP)^2+ (\delta \log \mbh)^2},
\end{equation}
where the error of $\log \mbh$ is $\delta \log \mbh=4.38\times \delta \log \sst$. $\delta \log \mbh\sim 0.19~ dex$ for 10\% percent uncertainty in  \sst. The error of four kinds of VP can be derived from,
\begin{equation}\label{eq5}
\begin{split}
\delta\vpfm &=\vpfm\sqrt{(\frac{\delta\tau}{\tau})^2+ (2\frac{\delta\fwm}{\fwm})^2},\\
\delta\vpsm &=\vpsm\sqrt{(\frac{\delta\tau}{\tau})^2+ (2\frac{\delta\sm}{\sm})^2},\\
\delta\vpfr &=\vpfm\sqrt{(\frac{\delta\tau}{\tau})^2+ (2\frac{\delta\fwr}{\fwr})^2},\\
\delta\vpsr &=\vpsr\sqrt{(\frac{\delta\tau}{\tau})^2+ (2\frac{\delta\sr}{\sr})^2},\\
\end{split}
\end{equation}
In Table \ref{tab3}, we present \mbh and four kinds of $f$ of low-z AGN RM sample for our calibration. For one source with multiple RM measurements, we calculate a mean value of $\log f$ to do the calibration. Considering a scaled factor of 3.80 for pseudobulges, we also give the values of $f$ in the bracket in Table \ref{tab3}.

In Fig. \ref{Fig1}, we show the distributions of four kinds of the factor $f$. The red line is the $f$ distribution for our total sample  excluding pseudobulges. The black line and the  blue dash-line are the $f$ distributions for our total sample adopting different \ms relation for pseudobulges with $\alpha=-0.51$ (PB unscaled $f$) and $\alpha=-1.09$ (PB scaled $f$ by a factor of 3.80), respectively. The mean value and the standard deviation of $f$ are shown in each panel (see also Table \ref{tab4}). Each kind of $f$ has a range of about two order of magnitude.

We separate the sample into Population 1 (Pop 1)  and Population 2 (Pop 2) following the suggestion by \cite{Co06}. Population 1 is \dhb(mean) $<$  2.35 and Population 2 is \dhb(mean) $\ge$2.35, where \dhb(mean) is measured from the mean spectrum. We also separate the sample into Population A (Pop A) and Population B (Pop B) \citep{Su00}, Population A is $\fwm < 4000~\kms$ and $\fwm \ge 4000~ \kms$. We calculate the average values of the factor $f$ for these populations for all sources, and all sources excluding pseudobulges shown in Table \ref{tab4}.


For different kinds of $f$, K-S tests are present in the last two lines in Table \ref{tab4} between Pop 1 and Pop 2, and between Pop A and Pop B. For Pop A and Pop B with a divided value of $\rm FWHM(\hb)=4000~ \kms$, the null hypothesises in K-S test that the data sets of $f_{\rm F, mean}$ are drawn from the same distribution $prob(\rm KS)$ is smaller than that of $f_{\rm \sigma, mean}$ for total sample or sample excluding pseudobulges (Values of $prob(\rm KS)<0.10$ are highlighted in boldface in Table \ref{tab4}). It is same for $f$ from the rms spectrum. About K-S test between Pop A and Pop B, for all AGNs with PB scaled by a factor of 3.80, $prob(\rm KS)$ is 0.08 for $f_{\rm F, mean}$ and  0.09 for $f_{\rm F, rms}$. Their significances are only higher than $1\sigma$, but weaker than $2\sigma$. For all AGNs with PB unscaled by a factor of 3.80, $prob(\rm KS)$ becomes smaller, 0.03 for $f_{\rm F, mean}$ and 0.01 for $f_{\rm F, rms}$. Their significances are higher than $2\sigma$, but weaker than $3\sigma$. For ellipticals and classical bulges, $prob(\rm KS)$ is 0.04 for $f_{\rm F, mean}$ and 0.01 for $f_{\rm F, rms}$, which are higher than $2\sigma$, but weaker than $3\sigma$. Therefore, these differences between Pop A and Pop B exist but not significant. For all AGNs with PB unscaled by a factor of 3.80, $prob(\rm KS)$ is 0.10 for $f_{\rm \sigma, mean}$ and 0.12 (about $1\sigma$) for $f_{\rm \sigma, rms}$. These values for $\sigma_{\rm \hb}$-based $f$ are larger than that for the corresponding $\rm FWHM(\hb)$ - based $f$. It implies that the FWHM-based $f$ has a relation with FWHM, while $\sigma$-based $f$ dose not has a relation with FWHM. It is consistent with the result by \cite{Co06}.




Considering $\sigma_{\rm \hb}$ from the mean/rms spectra for ellipticals and classical bulges, the average value of the factor of $\sigma_{\rm \hb}$-based $f$, $\log f_{\rm \sigma, mean}=0.76\pm 0.34$, and $\log f_{\rm \sigma, rms}=0.78\pm 0.37$. It is consistent with the results by others \citep[e.g.][]{HK14}. Including pseudobulges and the same \ms relation like classical bulges, the average value of the factor $f$ would be larger. Including pseudobulges and the \ms relation for pseudobulges suggested by \cite{HK14}, the average value of the factor $f$ would be smaller. Using $\rm FWHM(\hb)$ as the $\Delta V$ tracer from the mean/rms spectra for ellipticals and classical bulges, the average value of $\rm FWHM(\hb)$-based $f$, $\log f_{\rm F, mean}=0.08\pm 0.46$, and $\log f_{\rm F, rms}=0.17\pm 0.40$. They are about 4-5 times smaller than that derived from $\sigma_{\rm \hb}$. Including pseudobulges, they has the same trend like the case of $\sigma_{\rm \hb}$-based $f$.

\subsection{The relation of $f$ with observational parameters}

\begin{figure*}
\includegraphics[angle=-90,width=3.0in]{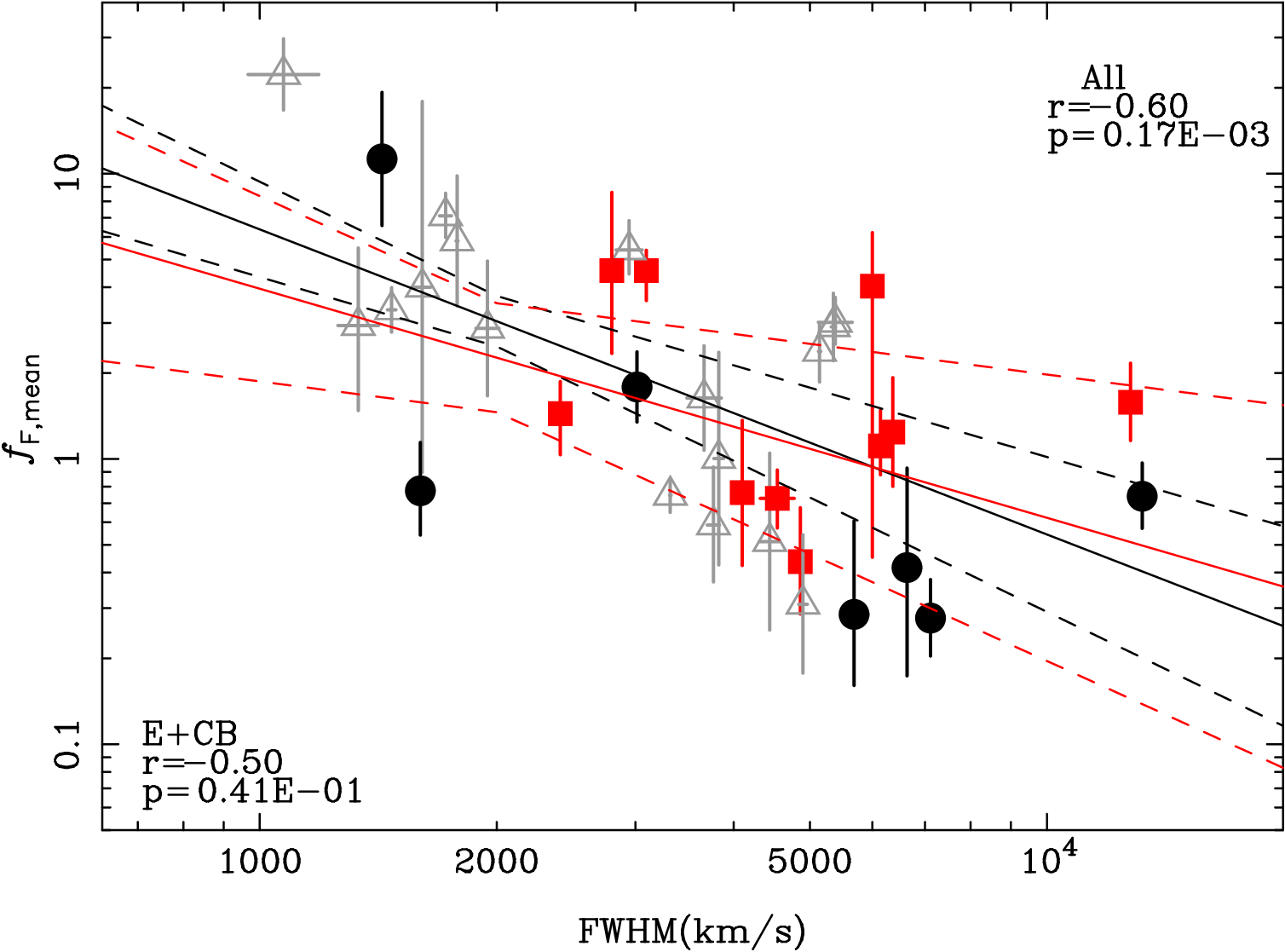}\hfill
\includegraphics[angle=-90,width=3.0in]{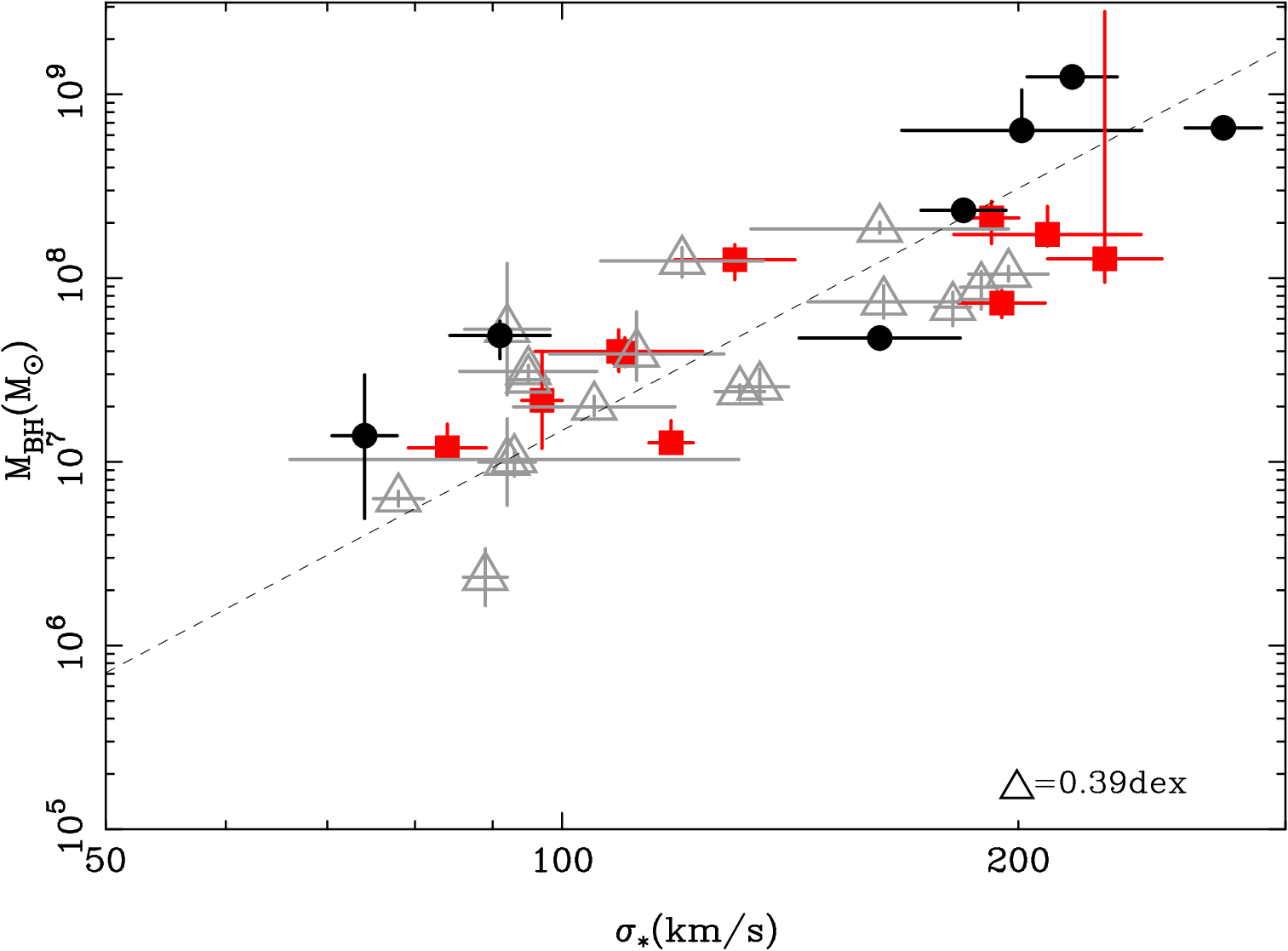}
\caption{Left: $f_{\rm F,mean}$ vs. \fwm. Black circles denote ellipticals. Red squares denote classical bulges. Gray triangles represent pseudobulges. The solid line is our best-fitting relation derived from BCES(Y$\vert$X), the dash lines represent the 1 $\sigma$ scatter range. The red lines are that only for 17 classical bulges and ellipticals. Right: the SMBH mass calculated based on the $f_{\rm F, mean}$ by Equation \ref{eq6} (the left panel) versus \sst, the dash line is the \ms relation \citep{KH13}.}
\label{Fig2}
\end{figure*}

\begin{figure*}
\includegraphics[angle=-90,width=3.0in]{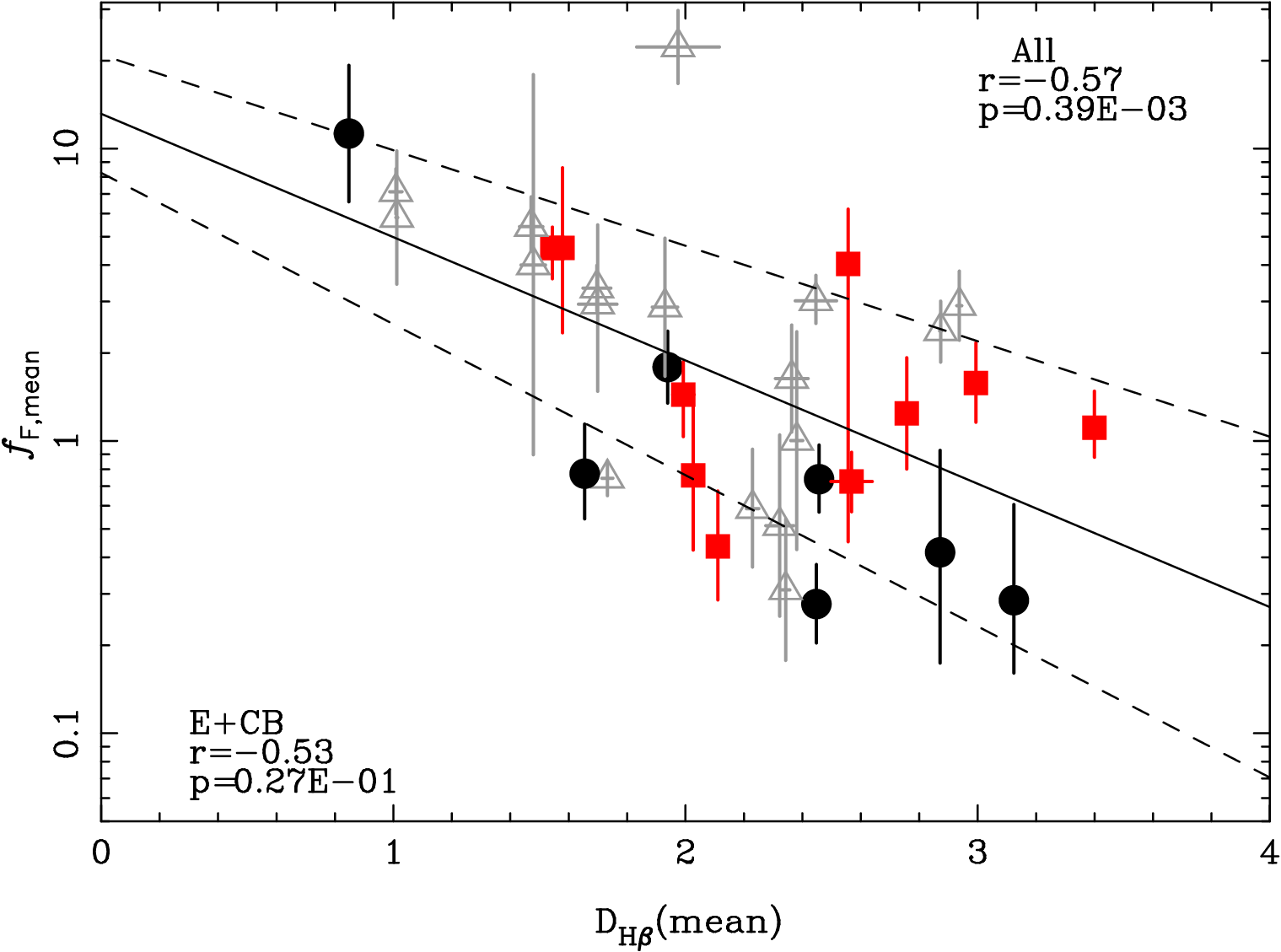}\hfill
\includegraphics[angle=-90,width=3.0in]{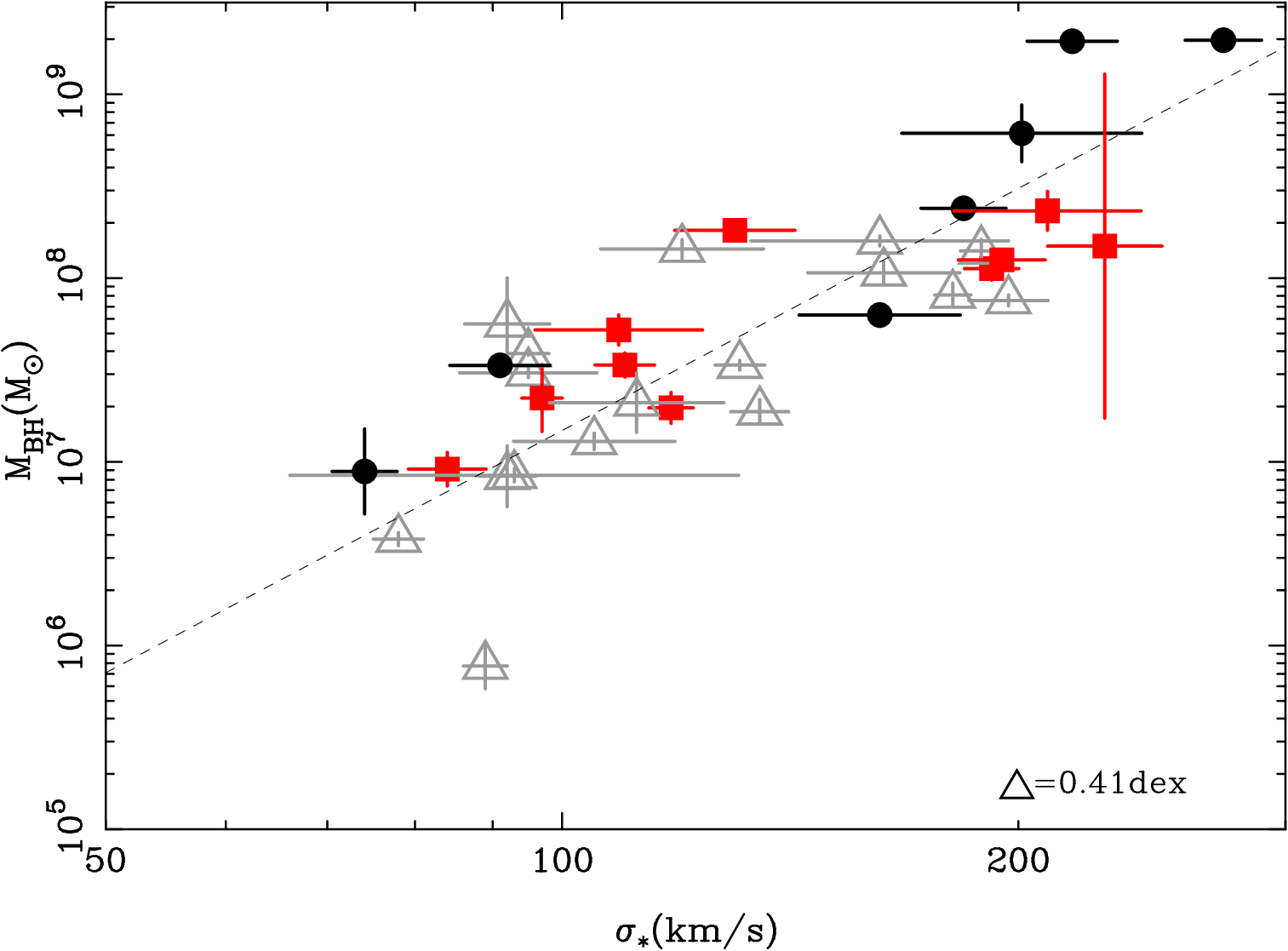}
\caption{Left: $f_{\rm F,mean}$ vs. \dhb.  Right: the SMBH mass versus \sst. The symbols and the lines are the same as Fig. \ref{Fig2}. }
\label{Fig3}
\end{figure*}


\begin{figure*}
\includegraphics[angle=-90,width=3.0in]{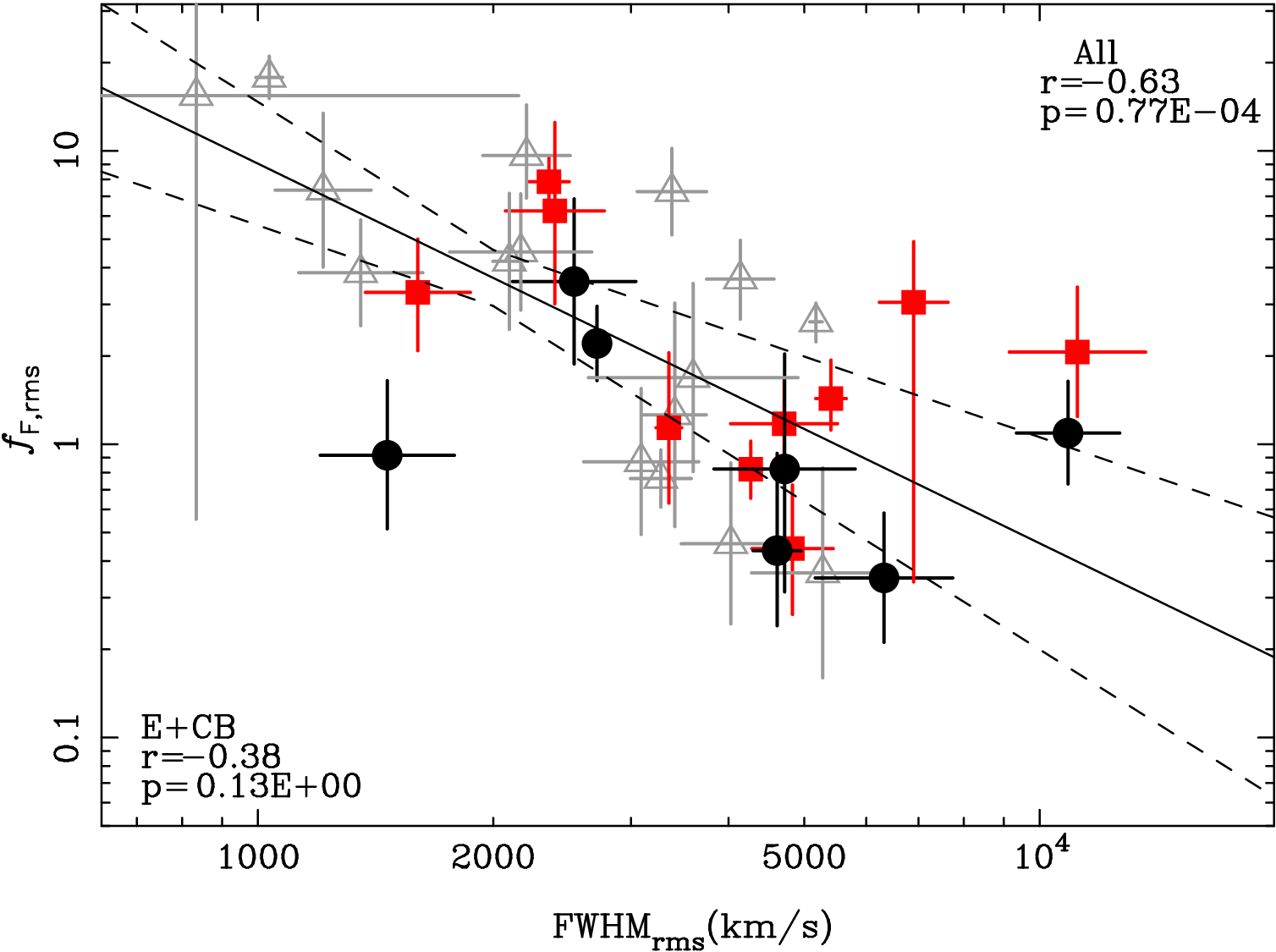}\hfill
\includegraphics[angle=-90,width=3.0in]{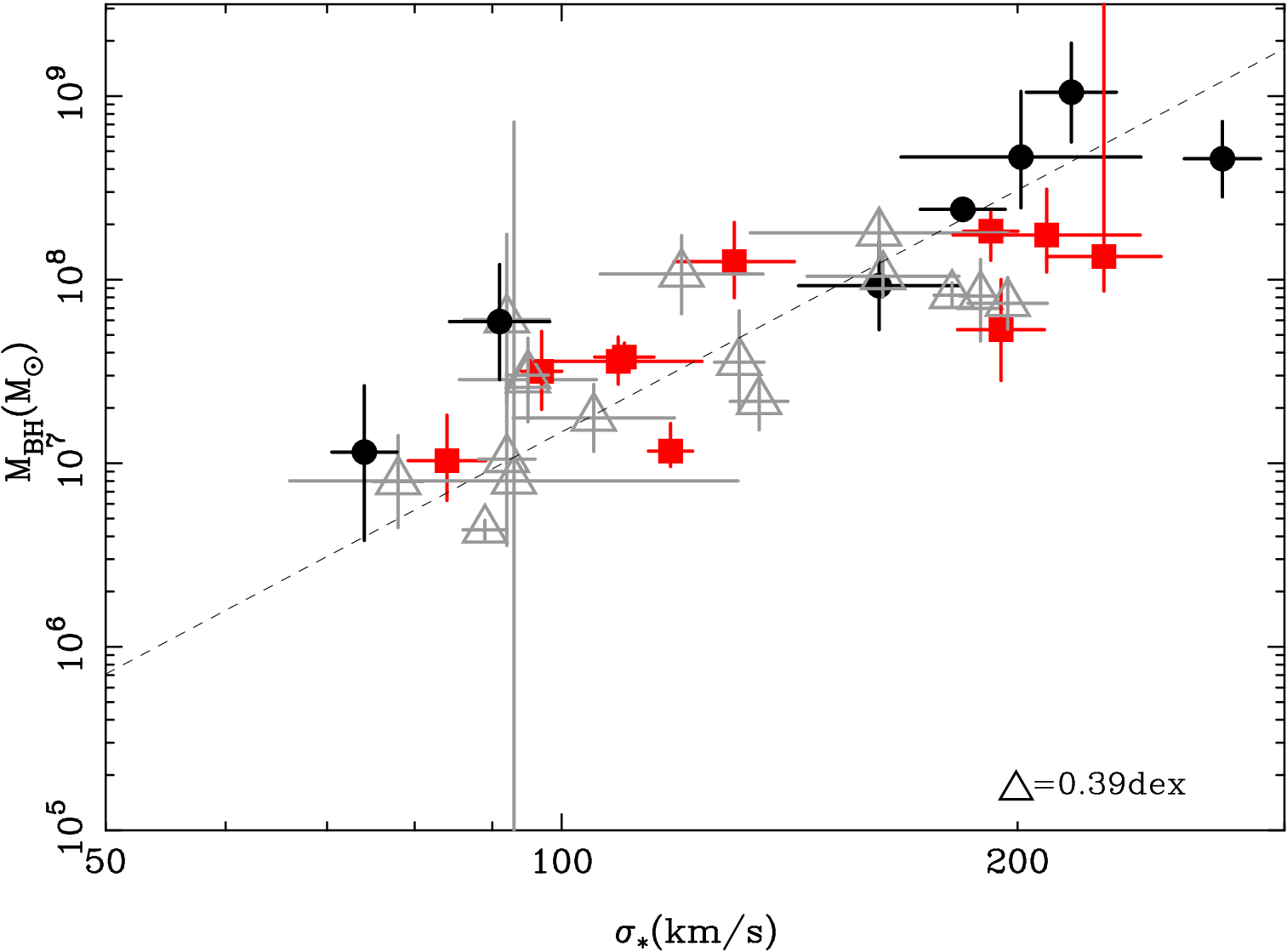}
\caption{Left: $f_{\rm F,rms}$ vs. \fwr.  Right: the SMBH mass versus \sst. The symbols and the lines are the same as Fig. \ref{Fig2}.  }
\label{Fig4}
\end{figure*}

In Table  \ref{tab4}, we notice that FWHM-based $f$ is indeed a bias quantity affected by \hb FWHM or \dhb but $\sigma_{\rm \hb}$-based $f$ is a less biased quantity, which is consistent with \cite{Co06}.
To investigate the relations between the virial factor $f$ and the observational parameters, we use the observational parameters of \sst, \fwm, \sm, \dhb(mean), \fwr, \sr, \dhb(rms), \rfe, $\tau$ and \lv. For pseudobulges, we use the same \ms relation like elliptical and classical bulges to calculate $f$, or a scaled \ms relation by \cite{HK14}. We calculate the Spearman correlation coefficient $R_{\rm s}$ and probability of the null hypothesis $p_{\rm null}$ between $f$ and these observational parameters, which are presented in Table \ref{tab5}. We highlight the correlations in boldface with the significant value ($|R_{\rm s}|>0.5$ or $p_{\rm null}<0.01$) in Table \ref{tab5}. If adopting different \ms relation for pseudobulges, the smaller intercept by a factor of 3.8 according to \cite{HK14}, it would lead to a smaller $f$ by a factor of 3.8 and the weakness of correlations with large correlation coefficients for total sample (Table \ref{tab5}). It was suggested that both active galaxies hosting pseudobulges and active galaxies with central bars may follow the same \ms relation of the elliptical/classical bulges \citep{Gr13,Wo15}. Therefore, we use the results from the total sample including pseudobulges with the  same \ms relation like ellipticals and classical bulges.


From Table \ref{tab5},  we can't find a significant correlation between $f$ and \sst ($R_s=0.23, 0.09, 0.09, 0.05$ for four kinds of $f$). For \sst-based \mbh, it suggests that there is no significant correlation between $f$ and \mbh, which is not consistent with the result by \cite{Williams2018}. We can't find a significant correlation between FWHM-based $f$ ($R_s=-0.03, -0.09$) and \lv or the Eddington ratio \lb/\ledd ($R_s=-0.13, -0.17$), which are consistent with the results by \cite{Williams2018}. It is found that there are correlations between $\sigma$-based $f$ and \lb/\ledd. We also find that there are some significant relationships between $f$ and the observational parameters. For $f_{\rm F, mean}$ and $f_{\rm F,rms}$, they have significant correlations with \fwm, \dhb(mean). However, for $f_{\rm \sigma, mean}$ and $f_{\rm \sigma,rms}$, they have no significant correlation (except with \sr, \lb/\ledd) considering their large $p_{\rm null}$. Excluding pseudobulges, a smaller number of sources weaken these significant correlations (see Table \ref{tab5}). In Table \ref{tab5}, $f_{\rm \sigma, rms}$ has no significant relation with \fwm, \dhb(mean), \sm, \rfe, and $R_{s}=-0.26, -0.12,-0.27, 0.35$, respectively. \cite{Wo10} also used a smaller sample to investigate the \mbh scatter relation with the Eddington ratio and line-shape measurements from the mean spectrum, which is similar to the relations of $f_{\rm \sigma, rms}$ with \fwm, \dhb(mean), \sm, \rfe. They can not find significant correlations, which is consistent with our results.


The left panel of Fig. \ref{Fig2} shows $f_{\rm F,mean}$ versus \fwm, where the same \ms relation like ellipticals and classical bulges is adopted for pseudobulges, i.e., unscaled $f$ for pseudobulges.  We find that, for all source in our sample, $f_{\rm F,mean}$ has a strong correlation with \fwm, $R_{\rm s}=-0.60$, $p_{\rm null} = 1.7\times10^{-3}$. We use the bivariate correlated errors and scatter method \cite[BCES;][]{AB96} to perform this linear regression. The BCES (Y$\vert$X) best-fitting relation for our total sample between $f_{\rm F, mean}$ and \fwm is,
\begin{equation}\label{eq6}
\log f_{\rm F,mean}=-(1.11 \pm 0.27)\log \frac{\fwm}{2000~ \kms} + (0.48 \pm 0.09),
\end{equation}
The relation is plotted as solid line in the left panel of Fig. \ref{Fig2}, i.e., $f_{\rm F, mean} \propto \fwm^{-1.11\pm 0.27}$. The dash line present $1\sigma$ scatter range. In Fig. \ref{Fig2}, it is found that AGNs with pseudobulges follow this relation. Excluding pseudobulge (gray triangles), i.e., only for ellipticals (black circles) and classical bulges (red squares), $R_{\rm s}=-0.50$, $p_{\rm null} = 0.041$. The weaker correlation is due to smaller number of AGNs excluding pseudobulges, and the BCES (Y$\vert$X) best-fitting relation is $\log f_{\rm F,mean}=-(0.80 \pm 0.45)\log \frac{\fwm}{2000~ \kms} + (0.36 \pm 0.19)$ (red lines in Fig. \ref{Fig2}). Adopting scaled $f$ for pseudobulges, the correlation becomes weaker for total sample (see Table \ref{tab5}). It is also consistent that the BLRs kinematics is independent to the morphology of galaxies, such as relative small bulges or bars. Considering that pseudobulges follow the $f_{\rm F,mean}-\fwm$ relation for classical bulges and ellipticals (see Fig. \ref{Fig2}), we use Equation \ref{eq6} as the $f$ calibration from the total sample with $f$ unscaled pseudobulges.  We also use the BCES (Y$\vert$X) to find the relations between $f_{\rm F, mean}$ and \dhb(mean), \fwr as follows,
\begin{equation}\label{eq7}
\begin{split}
& \log f_{\rm F, mean}=-(0.422 \pm 0.095) \dhb(\rm mean) + (1.12 \pm 0.20), \\
& \log f_{\rm F, rms}=-(1.29 \pm 0.38 )\log \frac{\fwr}{2000~ \kms} + (0.57 \pm 0.10),
\end{split}
\end{equation}

For each object, using the best-fitting relation between $f_{\rm F, mean}$ and \fwm (Equation \ref{eq6}), we can calculate the expected factor $f_{\rm F, mean}$ from the observational parameter of \fwm, and then calculate the calibrated \mbh from Equation \ref{eq2}. We select some strong correlations in Table \ref{tab5} to shown in Figs \ref{Fig2}, \ref{Fig3}, \ref{Fig4}. In the right panel of Fig. \ref{Fig2}, we show the calibrated \mbh from $f_{\rm F, mean}$ versus \sst, as well as the the relation by \cite{KH13} (dash line). With respect to this \ms relation, the \mbh scatter is 0.39 dex (shown in right corner in the right panel of Fig. \ref{Fig2}).
With respect to two other relations shown in Equations \ref{eq6}, the scatters are 0.41 dex and 0.39 dex, shown in right corners in the right panel of Figs. \ref{Fig3}, \ref{Fig4}. These scatters in \mbh are smaller than that assuming a constant $f=1.5$ suggested by \cite{HK14}, which is about 0.49 dex.

From Equation \ref{eq6}, $f_{\rm F, mean} \propto \fwm^{-1.11\pm 0.27}$. Combining with Equation \ref{eq2}, $\mbh=f \times VP\propto \fwm^{0.89}$. For the rms spectrum, $f_{\rm F, rms} \propto \fwr^{-1.29\pm 0.38}$. Combining with Equation \ref{eq2},  $\mbh=f \times VP\propto \fwr^{0.71}$. The index of $0.89$, or $0.71$ is deviated from usually assumed value of $2$. It is consistent with the result by \cite{Pa12}, although their index is $1.73$. They suggested a FWHM or $\sigma_{\rm line}$ dependent virial factor $f$ to correct for the systematic difference of the geometry and kinematics of the gas contributing to the single-epoch line profile and that contributing to the rms spectra.
Recently, based on accretion disk model to fit SED, it was suggested that $f$ has a relation with FWHM(\hb) and the slope is $-1.17$, which is almost the same as ours \citep{Mejia2018}.
For NGC 5548, we also find a medium-strong  correlation between $f_{\rm F, mean}$ and $\fwm$ with $R_{\rm s}=-0.56, p_{\rm null}=0.02$ \citep{Lu2016}. The BCES (Y$\vert$X) best-fitting gives $f_{\rm F, mean} = (\fwm/2000~ km s^{-1})^{-0.97\pm 0.42}+(0.88\pm 0.21)$. Although it is consistent with that for all RM AGNs, the result of NGC 5548 means that the variation of an individual object could contribute some scatter to the relation between $f_{\rm F, mean}$ and \fwm. These results imply the effect of inclination in BLRs geometry \citep[e.g.][]{Sh14}.

For narrow line Seyfert 1 galaxies (NLS1s), smaller $\rm FWHM(\hb) < 2000~ \kms$ would lead to a larger $f_{\rm F, mean}$ than the usual adopted value of $\sim 1$ (Equation \ref{eq6}). Adopting $\fwm < 2000~ \kms$ in our sample of 34 RM AGNs, there are 9 NLS1s (Fig. \ref{Fig2}). 7 out of 9 NLS1s are pseudobulges with $f_{\rm F, mean} \sim 3-20$. It was found that NLS1s have smaller $R_{\rm BLR}$ expected than that from the empirical $R_{\rm BLR}-L$ relation \citep{Du18}. It is needed to consider the influence of these two contributions on \mbh calculation in NLS1s. On the other hand, for AGNs with very broad double-peaked \hb, their large \fwm would lead to a smaller $f_{\rm F, mean}$. We also notice that there are two AGNs with $\fwm > 10000~ \kms$ with the usual adopted value of $f \sim 1$. Considering a smaller number of these special AGNs, a large sample of RM AGNs of double-peaked AGNs and NLS1s  is needed for further investigation \citep{Bian2007, Bian2008, Wang2013}.

\section{Discussion}

\subsection{Cumulative fraction of $f$ and the structure of BLRs}
\begin{figure*}
\includegraphics[angle=-90,width=3.0in]{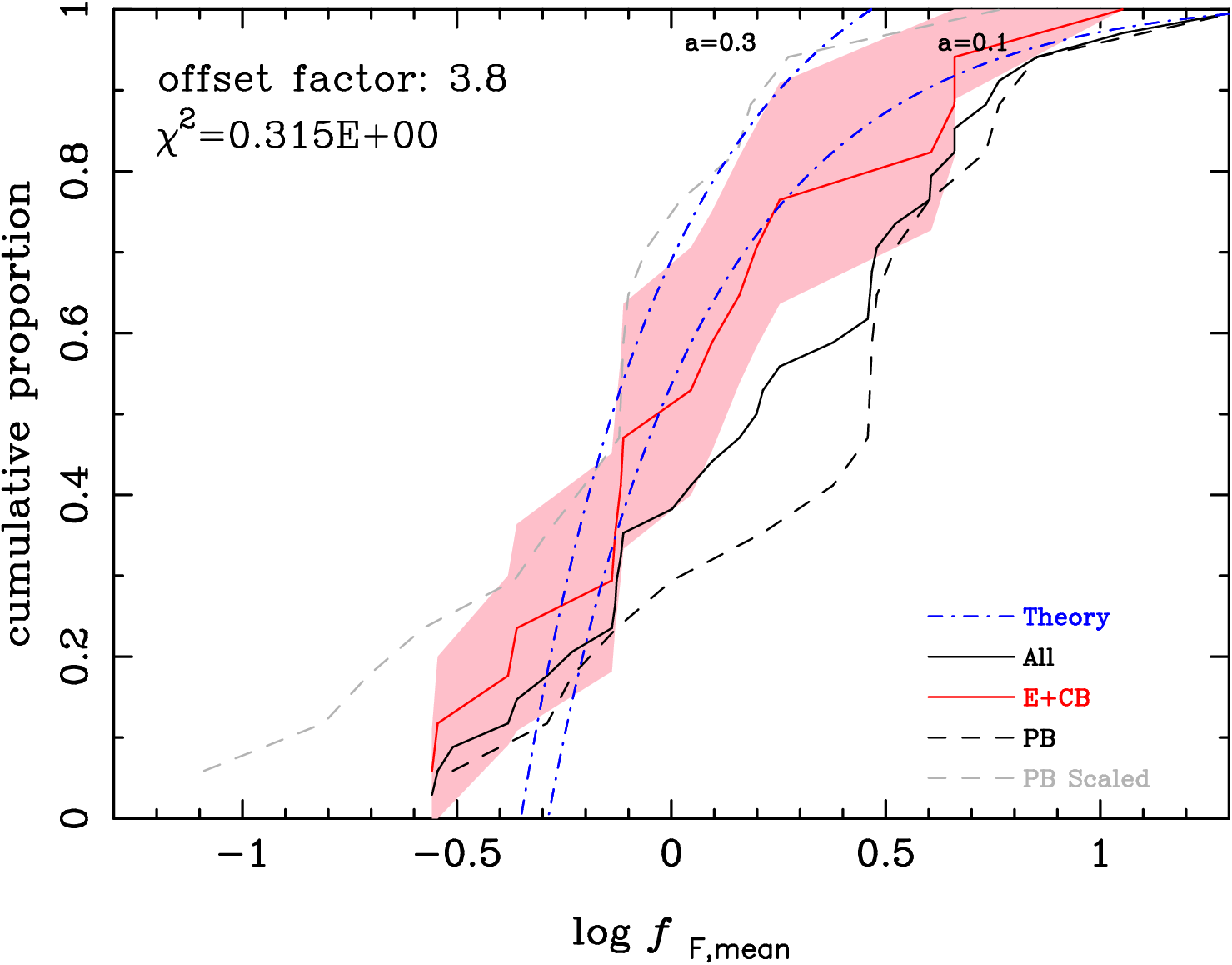}\hfill
\includegraphics[angle=-90,width=3.0in]{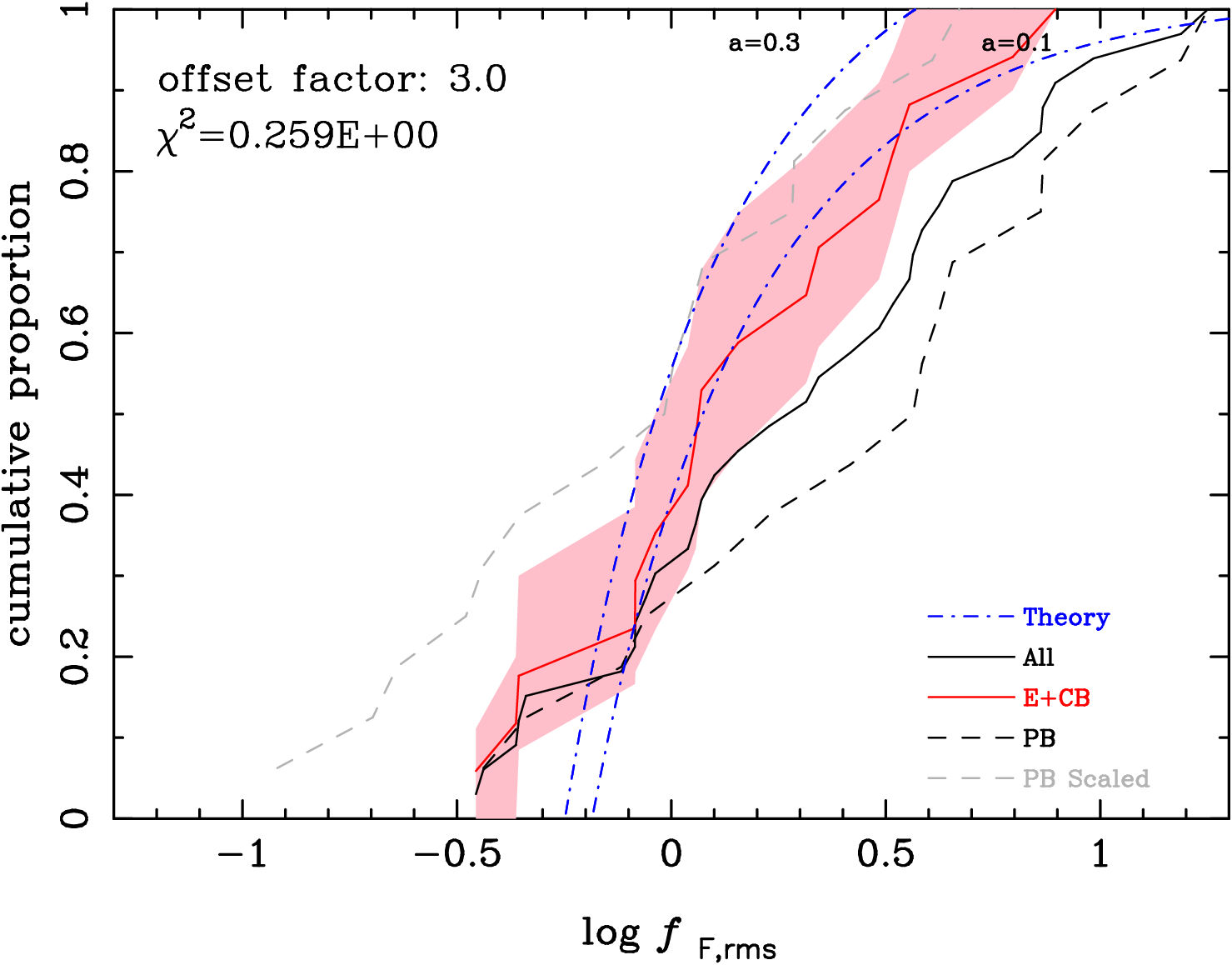}
\includegraphics[angle=-90,width=3.0in]{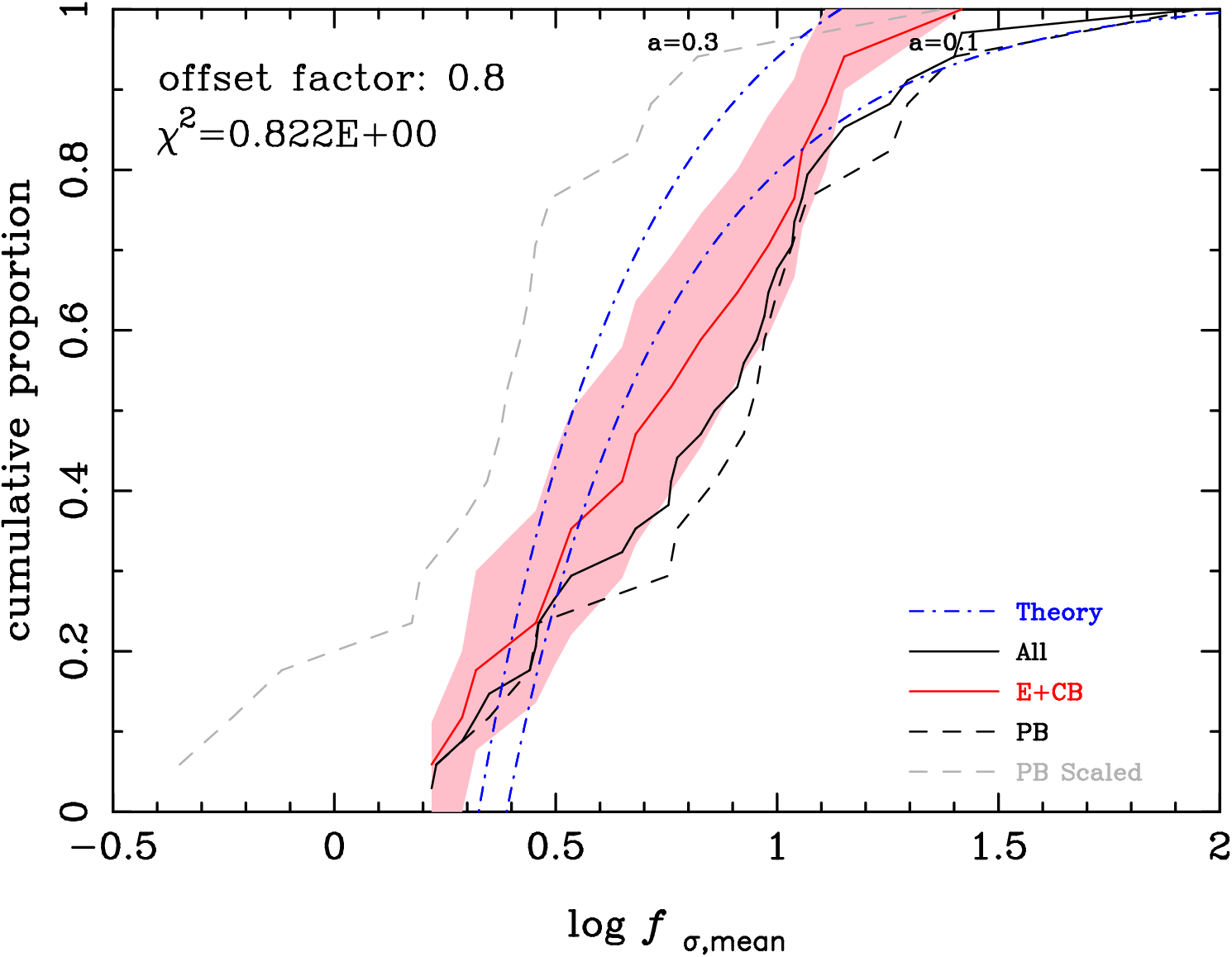}\hfill
\includegraphics[angle=-90,width=3.0in]{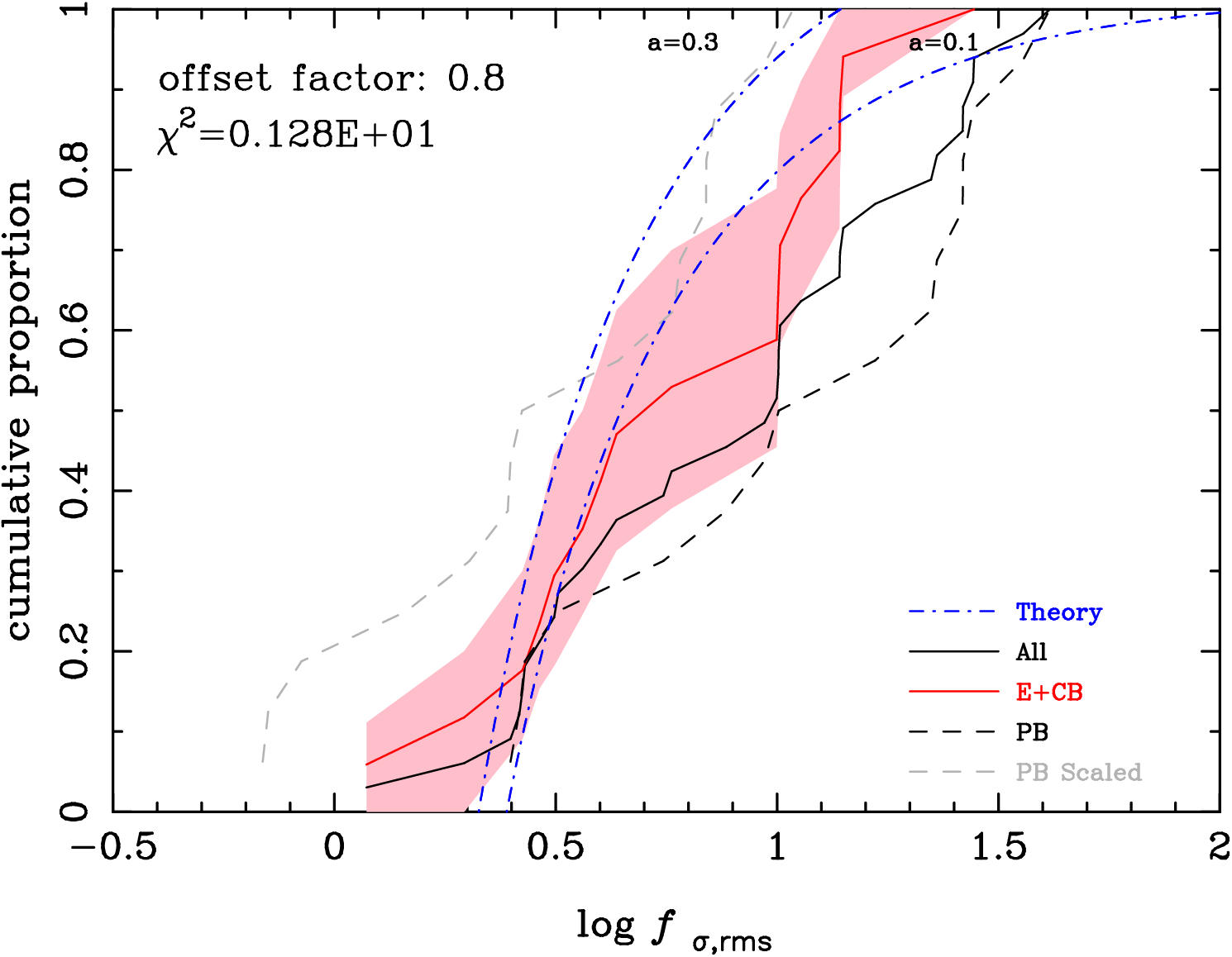}
\caption{Cumulative fraction of values of four kind of $f$ (top left: $f_{\rm F, mean}$; top right $f_{\rm F, rms}$; bottom left: $f_{\rm \sigma, mean}$, bottom right: $f_{\rm \sigma, rms}$), compared to the theoretical distributions (blue dot dash lines) for two thick-disk models of a=0.1 and a=0.3. The “offset factor” is the value by which the theoretical value has been divided to aid in comparison of the two distributions. The black line is for the total sample. The red line is for classical bulge and elliptical. The dash line is for pseudobulge. The gray dash line is for pseudobulge scaled with a factor of 3.8 \citep{HK14}. The bootstrap method is used to obtain the confidence regions (90\%) of the red line for classical bulges and ellipticals, which is shown as the pink area.}
\label{Fig5}
\end{figure*}

Fig. \ref{Fig5} shows the cumulative fraction of values of four kind of $f$ (top left: $f_{\rm F, mean}$; top right $f_{\rm F, rms}$; bottom left: $f_{\rm \sigma, mean}$, bottom right: $f_{\rm \sigma, rms}$). The black line is for the total sample. The red line is for classical bulge and elliptical. The bootstrap method is used to obtain the confidence regions (90\%) of the red line, which is shown as the pink area. The dash line is for pseudobulge. The gray dash line is for pseudobulge scaled with a factor of 3.8 suggested by \cite{HK14}. The cumulative fraction of $f$ could be tested by BLRs models \citep{Co06}.

Assuming that BLRs are made of two dynamically distinct component, i.e., a disk and a wind, the observational velocity $v_{\rm obs}$  could be expressed as \citep{Co06},
\begin{equation}\label{eq8}
v_{\rm obs}=[k_1^2(a^2+sin^2\theta)V_{\rm Kep}^2+k_2^2V^2_{\rm out}cos^2\theta]^{1/2},
\end{equation}
where $V_{\rm out}$ is the outflow velocity, assumed to be normal to the disk, and $k_1,k_2$ are the contributions of the thick disk and of  the wind, respectively, $a$ is the ratio of the scale height of the thick disk to the radius R, or the ratio of the turbulent velocity to the local Keplerian velocity $V_{\rm Kep}$ at the radius R, $\theta$ is the inclination of the thick-disk of BLRs to the line of sight. For a simple model of thick-disk BLRs, we neglect the contribution of outflow in \hb profile, and $v_{\rm obs}=(a^2+sin^2\theta)^{1/2}V_{\rm Kep}$.
Considering the effect of inclination $\theta$ in this model of thick-disk BLRs, the $f$ is,
\begin{equation}
f^{\rm disk}=\frac{R_{\rm BLR} V_{\rm Kep}^2/G}{R_{\rm BLR} v_{\rm obs}^2/G}=\frac{V_{\rm Kep}^2}{v_{\rm obs}^2}=\frac{1}{a^2+sin^2\theta},
\end{equation}
According to the unified scheme \citep{AM85}, type I AGNs can be observed for a inclination less than $\theta_0$. We adopt $\theta_0=45^o$, $a=0.1$ or $0.3$ \citep{Co06}. $\theta$ can be calculated for a value of $f^{\rm disk}$. The
probability of seeing an object at an inclination angle $\theta$ per unit angle interval is $sin\theta/(1-cos\theta_0)$. And the theoretical cumulative fraction of $f^{\rm disk}$ (i.e. at an inclination angle $\theta$) can be calculated from the integral of solid angle from this $\theta$ to $\theta=0^o$. The theoretical lines of the cumulative fraction from the thick-disk BLRs are shown as blue dot-dashed lines in Fig. \ref{Fig5}.

Considering the difference between $v_{\rm obs}$ and $\Delta V$ in Equation \ref{eq2}, there is an "offset factor" (shown in these figures)  by which the theoretical $f^{\rm disk}$ has been divided to aid in comparison of the theoretical distribution of $f^{\rm disk}$ and the observed distribution $f^{\rm obs}$ \citep{Co06}. The best fit is reached by minimizing $\chi^2$, where 1$\sigma$ error from bootstrap method is adopted. The values of the "offset factor" in the fitting are shown in Fig. \ref{Fig5}.  Adopting FWHM(\hb) as the tracer of $v_{\rm obs}$ (top two panels in Fig. \ref{Fig5}), for our subsample of 17 classical bulges and ellipticals, it is found that the cumulative fractions of $f_{\rm F, mean}$ and $f_{\rm F, rms}$ are consistent with the theoretical line ($a=0.1$) for large values of $f$, $\chi^2=0.353, 0.261$. For $a=0.3$, $\chi^2=4.16, 3.10$, which imply that larger turbulence velocity is not required to model the fraction distributions of $f_{\rm F, mean}$ or $f_{\rm F, rms}$. For $f$ unscaled or scaled pseudobulge, this model of thick-disk BLRs has a larger fraction difference ($\chi^2=2.64$ or $0.85$ for $f_{\rm F, mean}$ and $\chi^2=2.56$ or $2.06$ for $f_{\rm F, rms}$). Using $\sigma_{\rm \hb}$ to trace $v_{\rm obs}$ for our subsample of 17 classical bulges and ellipticals (bottom two panels in Fig. \ref{Fig5}), the theoretical line ($a=0.1$) has a larger fraction difference ($\chi^2=0.798, 1.21$). These results imply that FWHM($\rm \hb$) has some dependence on the inclination, while $\sigma_{\rm \hb}$ is insensitive to the inclination. It is consistent with the result by \cite{Co06}. For pseudobulge, FWHM($\rm \hb$) or $\sigma_{\rm \hb}$ seems insensitive to the inclination.

\subsection{Comparison of $f$ with the BLRs dynamical model}
\begin{figure}
\includegraphics[angle=-90,width=3.5in]{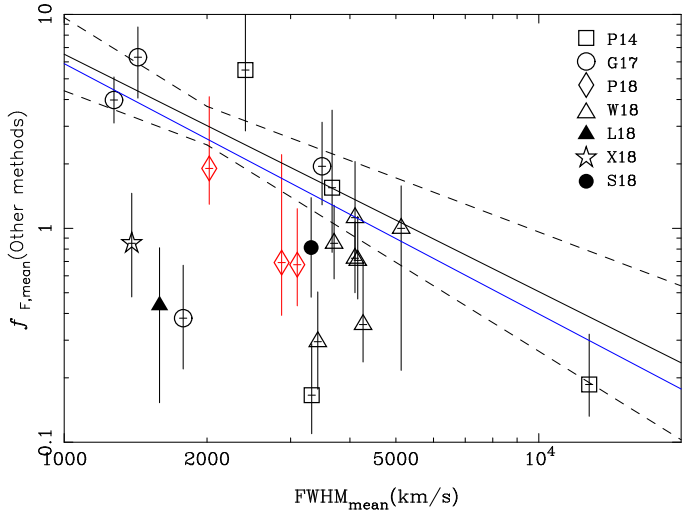}
\caption{$f_{\rm F, mean}^{\rm Model}$ versus $\rm FWHM_{\rm mean}$ of \hb, where $f_{\rm F, mean}^{\rm model}$ is calculated from the BLRs dynamical modelling of the continuum light curve and the \hb line profiles, X-ray variability and resolved Pa$\alpha$ emission region. The solid line is our calibration correlation of Equation \ref{eq6} and the dash lines show the range of 1 $\sigma$ uncertainty.The blue line is the relation from the method by the standard accretion disk model fit of AGNs SED \citep{Mejia2018}. Different symbols denote different references (see Table \ref{tab6}).  }
\label{Fig6}
\end{figure}
The cross-correlation and $R_{\rm BLR}-L$ relation yield the BLRs size, but they do not provide information about the gas structure or dynamics, which is needed to determine $f$ for an individual AGN. Recent efforts have aimed to measure $f$ by modelling the structure and dynamics of the BLR directly. Using BLRs dynamical models to fit simultaneously the variations in the \hb flux and the detailed shape of the \hb profile, the $f_{\rm F, mean}^{\rm Model}$ has been derived for 17 RM AGNs \citep{Williams2018,Li2018}. \cite{Pan2018} measured the \mbh of RM AGN 1H 0323+342 through the X-ray variability. \cite{Sturm2018} used VLTI to resolve the emission region of Pa$\alpha$ and measured the \mbh in 3C 273. It suggested that $\log f_{\rm F, mean} = -0.09\pm 0.23$. In Table \ref{tab6},  the $f_{\rm F, mean}^{\rm Model}$ from BLRs dynamical model, X-ray Variability, resolved Pa$\alpha$ emission region for 19 RM AGNs are presented. In Fig. \ref{Fig6}, we show the $f_{\rm F, mean}^{\rm Model}$ versus \hb $\rm FWHM_{\rm mean}$ for these 19 RM AGNs. For comparison, we also show our calibrated correlation of $f_{\rm F, mean}$ with $\rm FWHM_{\rm mean}$ as a solid line. The blue line is the relation from the method by the standard accretion disk model fit of AGNs SED \citep{Mejia2018}, which is consistent well with our relation shown as black solid line. Different symbols denote different references. In Fig. \ref{Fig6}, P14, G17, P18, W18, L18 and X18 S18 are \cite{Pancoast2014,Gr17a, Williams2018, Pancoast2018, Li2018, Pan2018, Sturm2018}, respectively. It is found that 12 out of 19 RM AGNs follow our calibrated correlation with 1 $\sigma$ uncertainty.

\subsection{The \ms relation for High-z AGNs}
\begin{figure}
\includegraphics[angle=-90,width=3.5in]{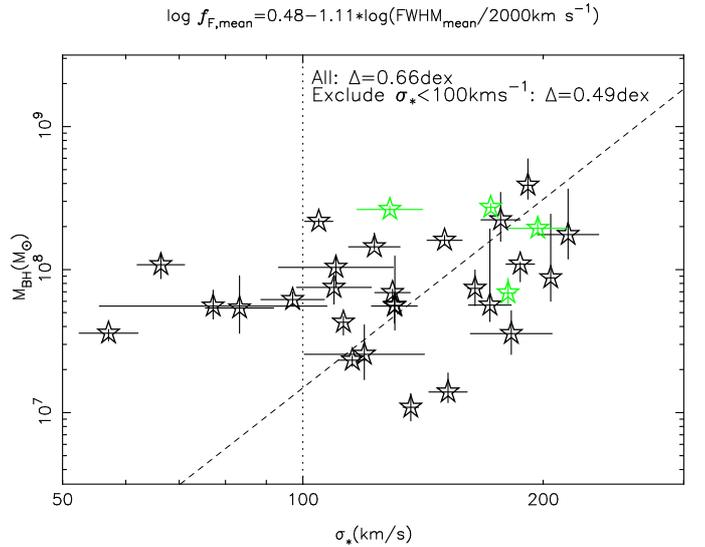}
\caption{The SMBH mass calculated based on the $f_{\rm F, mean}$ by Equation \ref{eq6} versus \sst for 30 SDSS RM AGNs, the dash line is the \ms relation for quiescent galaxies \citep{KH13}. The four green stars denote AGNs with \ha lags. The vertical dot line is \sst=100 \kms.}
\label{Fig7}
\end{figure}

Like the method to calculate the \mbh scatter based on the calibration of $f$ from the line width, i.e., the best-fitting relation between $f_{\rm F, mean}$ and \fwm (Equation \ref{eq6}), we can calculate \mbh for a high-z sample of 30 AGNs ($z~\sim 0.1 - 1.0$) from SDSS RM Project \citep{Gr17b}. In Fig. \ref{Fig7}, \mbh scatter is 0.66 dex. These scatters in \mbh are smaller than that assuming a constant $f=1.5$ suggested by \cite{HK14}, which is about 0.75 dex. For SDSS RM AGNs with \sst lower than 100\kms, they deviate the \ms relation and locate above the \ms relation. For AGNs with \sst larger than 100 \kms, they follow \ms relation by \cite{KH13}, but with a \mbh scatter of 0.49 dex. It is smaller than the scatter of 0.64 dex assuming a constant $f=1.5$ suggested by \cite{HK14}. For the total sample of high-z SDSS RM AGNs, the relation between \mbh and \sst is flatter than that by \cite{KH13}. It is possibly due to their un-coevolution between the SMBH and the bulge, or the large systematic uncertainties in \mbh and \sst. A larger sample with better measurements of \mbh and \sst is needed in the future.

\section{Conclusions}

Using a sample of 34 RM AGNs with measured bulge stellar velocity dispersions \sst, we calculate the virial coefficient $f$ based on the \ms relation for quiescent galaxies suggested by \cite{KH13}. For our sample of these 34 RM AGNs, there are 8 classified as ellipticals, 9 classified as classical bulges, 17 classified as pseudobulges. The main conclusions can be summarized as follows:
\begin{itemize}
\item Considering four tracers of the velocity of the BLRs clouds, i.e., $\rm FWHM(\hb)$ or $\sigma_{\rm \hb}$ from the mean or rms spectrum, there are four kinds of the factor $f$. Each kind of $f$ has a range of about two order of magnitude. For K-S test between Pop A and Pop B, $prob(\rm KS)$ for $\rm FWHM(\hb)$-based  $f$ is smaller than for the corresponding $\sigma_{\rm \hb}$-based $f$. For $\sigma_{\rm \hb}$-based $f$, the difference between Pop A and  Pop B is smaller than for $\rm FWHM(\hb)$-based  $f$. 

\item Significant correlations are found between FWHM-based $f$ and some observational parameters, e.g, $\rm FWHM (\hb)$, $\sigma_{\rm \hb}$, \dhb. However, for $\sigma_{\rm \hb}$-based $f$, these relations disappear or become weaker. For our sample, $f_{\rm F, mean} \propto \fwm^{-1.11\pm 0.27}$. For a single object NGC 5548, a similar strong correlation between $f_{\rm F, mean}$ and \fwm is also found. These strong correlations suggest a dependence of $f$ on FWHM or $\sigma_{\rm line}$ and imply the effect of inclination of BLRs to our line of sight. These strong correlations can be used to calibrate the factor $f$. When using the FWHM from the single-epoch spectrum to calculate the \mbh, the variable FWHM-based $f$ should be considered, especially for NLS1s and double-peaked AGNs.

\item Using a simple model of thick-disk BLRs, we calculate the theoretical cumulative fraction of $f$ and compare with the distributions of four kinds of $f$. It is found that the cumulative fractions of $f_{\rm F, mean}$ and $f_{\rm F, rms}$ are consistent with the theoretical line ($a=0.1$) only for the classical bulge and ellipticals, where FWHM is used as the tracer of $v_{\rm obs}$. This results imply that, as the tracer of BLRs velocity, FWHM(\hb) has some dependence on the BLRs inclination, while line $\sigma_{\rm \hb}$ is insensitive to the inclination. For pseudobulges, FWHM(\hb) or line $\sigma_{\rm \hb}$ seems insensitive to the inclination. For a sample of 19 RM AGNs, BLRs dynamical model and other independent methods gave their factor $f$ in the SMBH mass calculation. We find that 12 out of these 19 AGNs follow our calibrated correlation between $f$ and $\rm FWHM_{\rm mean}$.

\item Based on the $\rm FWHM(\hb)$-based $f$ from the mean spectrum, the \mbh scatter is 0.39 dex. For a high redshift sample of 30 SDSS RM AGNs with measured \sst, we find that the \mbh scatter is larger than that for our sample of 34 low redshift RM AGNs. It implies the possibility of evolution of the \ms relation for high-z AGNs.

\end{itemize}

\section*{Acknowledgements}
We are very grateful to the anonymous referee for her/his instructive comments which significantly improved the content of the paper. This work is supported by the National Key Research and Development
Program of China (No. 2017YFA0402703). This work has been supported
by the National Science Foundations of China (Nos. 11373024,
11233003, and 11873032).

\begin{table*}
\caption{The low-z RM AGNs sample with measured \sst for the calibration of $f$.  }
\begin{lrbox}{\tablebox}
\begin{tabular}{lcllllllllllllll}
\hline
Name &Alternate&\rfe&\sst&Bulge&\lv &$\tau$&\fwm&\fwr&\sm&\sr&\vpfm&\vpsm& \vpfr &\vpsr&Ref.\\
&name&&\kms&Type&\ergs&days&\kms&\kms&\kms&\kms&$10^6\msun$&$10^6\msun$&$10^6 \msun$&$10^6\msun$\\
 (1)&(2)&(3)&(4)&(5)&(6)&(7)&(8)&(9)&(10)&(11)&(12)&(13)&(14)&(15)&(16)\\
\hline
3C  120&$\cdots$& 0.39&$162\pm20$&E/CB &$ 43.96\pm 0.06$&$ 27.2^{+  1.1}_{-  1.1}$&$ 1430\pm   16$&$ 2539\pm  466$&$ 1687\pm    4$&$ 1514\pm   65$&$  10.9^{+  0.5}_{-  0.5}$&$  15.1^{+  0.6}_{-  0.6}$&$  34.2^{+ 12.6}_{- 12.6}$&$  12.2^{+  1.2}_{-  1.2}$&1,2 \\
3C  390.3&$\cdots$& 0.14&$273\pm16$&E &$ 44.36\pm 0.03$&$ 47.9^{+  2.4}_{-  4.2}$&$13211\pm   28$&$10872\pm 1670$&$ 5377\pm   37$&$ 5455\pm  278$&$1631.5^{+ 82.0}_{-143.2}$&$ 270.3^{+ 14.0}_{- 24.0}$&$1105.0^{+344.0}_{-353.0}$&$ 278.2^{+ 31.6}_{- 37.4}$&1,2 \\
Ark 120&Mrk 1095& 0.83&$192\pm 8$&CB&$43.92\pm 0.06$&$ 34.7^{+  6.6}_{-  8.9}$&$ 6042\pm   35$&$ 5536\pm  297$&$ 1753\pm    6$&$ 1959\pm  109$&$ 247.2^{+ 47.1}_{- 63.5}$&$  20.8^{+  4.0}_{-  5.3}$&$ 207.5^{+ 45.3}_{- 57.7}$&$  26.0^{+  5.7}_{-  7.3}$&1,2 \\
             &&     &          &  &$43.57\pm 0.10$&$ 28.8^{+  3.2}_{-  5.7}$&$ 6246\pm   78$&$ 5284\pm  203$&$ 1862\pm   13$&$ 1884\pm   48$&$ 219.3^{+ 25.0}_{- 43.7}$&$  19.5^{+  2.2}_{-  3.9}$&$ 156.9^{+ 21.2}_{- 33.3}$&$  20.0^{+  2.4}_{-  4.1}$&1,2 \\
Arp 151&Mrk 40& 0.32&$118\pm 4$&CB&$ 42.50\pm 0.11$&$  3.6^{+  0.7}_{-  0.2}$&$ 3098\pm   69$&$ 2357\pm  142$&$ 2006\pm   24$&$ 1252\pm   46$&$   6.7^{+  1.3}_{-  0.5}$&$   2.8^{+  0.5}_{-  0.2}$&$   3.9^{+  0.9}_{-  0.5}$&$   1.1^{+  0.2}_{-  0.1}$&1,2 \\
Mrk 50 &$\cdots$&$\cdots$&$109\pm14$&CB&$ 42.83\pm 0.06$&$  8.7^{+  1.6a}_{-  1.5}$&$ 4101\pm   56^b$&$ 3355\pm  128^b$&$ 2024\pm   31^b$&$ 2020\pm  103^b$&$  28.4^{+  5.4}_{-  5.0}$&$   6.9^{+  1.3}_{-  1.2}$&$  19.0^{+  3.9}_{-  3.6}$&$   6.9^{+  1.5}_{-  1.4}$&1,3,4\\
Mrk 79&UGC 3973& 0.33&$130\pm12$&CB&$ 43.57\pm 0.07$&$ 25.0^{+  2.8}_{- 14.1}$&$ 5056\pm   85$&$ 5086\pm 1436$&$ 2314\pm   23$&$ 2137\pm  375$&$ 124.7^{+ 14.6}_{- 70.5}$&$  26.1^{+  3.0}_{- 14.7}$&$ 126.2^{+ 72.7}_{-100.7}$&$  22.3^{+  8.2}_{- 14.8}$&1,2 \\
             &&     &          &  &$ 43.67\pm 0.07$&$ 30.2^{+  1.4}_{-  2.1}$&$ 4760\pm   31$&$ 4219\pm  262$&$ 2281\pm   26$&$ 1683\pm   72$&$ 133.5^{+  6.4}_{-  9.4}$&$  30.7^{+  1.6}_{-  2.2}$&$ 104.9^{+ 13.9}_{- 14.9}$&$  16.7^{+  1.6}_{-  1.8}$&1,2 \\
             &&     &          &  &$ 43.60\pm 0.07$&$ 16.8^{+  7.1}_{-  2.2}$&$ 4766\pm   71$&$ 5251\pm  533$&$ 2312\pm   21$&$ 1854\pm   72$&$  74.5^{+ 31.6}_{- 10.0}$&$  17.5^{+  7.4}_{-  2.3}$&$  90.4^{+ 42.4}_{- 21.8}$&$  11.3^{+  4.9}_{-  1.7}$&1,2 \\
Mrk 202&$\cdots$& 0.57&$ 78\pm 3$&PB&$ 42.21\pm 0.18$&$  3.5^{+  0.1}_{-  0.1}$&$ 1471\pm   18$&$ 1354\pm  250$&$  867\pm   40$&$  659\pm   65$&$   1.5^{+  0.1}_{-  0.1}$&$   0.5^{+  0.04}_{-  0.04}$&$   1.3^{+  0.5}_{-  0.5}$&$   0.3^{+  0.1}_{-  0.1}$&1,2 \\
Mrk 509&$\cdots$& 0.13&$184\pm12$&E &$ 44.13\pm 0.05$&$ 67.6^{+  0.3}_{-  0.3}$&$ 3015\pm    2$&$ 2715\pm  101$&$ 1555\pm    7$&$ 1276\pm   28$&$ 119.9^{+  0.6}_{-  0.6}$&$  31.9^{+  0.3}_{-  0.3}$&$  97.2^{+  7.2}_{-  7.2}$&$  21.5^{+  0.9}_{-  0.9}$&1,2 \\
Mrk 590&NGC 863& 0.45&$189\pm 6$&PB&$ 43.07\pm 0.11$&$ 19.0^{+  1.9}_{-  3.9}$&$ 3729\pm  426$&$ 2566\pm  106$&$ 2169\pm   30$&$ 1935\pm   52$&$  51.6^{+ 12.9}_{- 15.8}$&$  17.4^{+  1.8}_{-  3.6}$&$  24.4^{+  3.2}_{-  5.4}$&$  13.9^{+  1.6}_{-  2.9}$&1,2 \\
             & &    &          &  &$ 43.32\pm 0.08$&$ 31.8^{+  3.4}_{-  8.6}$&$ 2744\pm   79$&$ 2115\pm  575$&$ 1967\pm   19$&$ 1251\pm   72$&$  46.7^{+  5.7}_{- 12.9}$&$  24.0^{+  2.6}_{-  6.5}$&$  27.8^{+ 15.4}_{- 16.9}$&$   9.7^{+  1.5}_{-  2.9}$&1,2 \\
             & &    &          &  &$ 43.59\pm 0.06$&$ 30.1^{+  2.4}_{-  2.3}$&$ 2500\pm   43$&$ 1979\pm  386$&$ 1880\pm   19$&$ 1201\pm  130$&$  36.7^{+  3.2}_{-  3.1}$&$  20.8^{+  1.7}_{-  1.6}$&$  23.0^{+  9.2}_{-  9.1}$&$   8.5^{+  2.0}_{-  2.0}$&1,2 \\
Mrk 1310&$\cdots$& 0.46&$ 84\pm 5$&CB&$ 42.28\pm 0.17$&$  4.2^{+  0.9}_{-  0.1}$&$ 2409\pm   24$&$ 1602\pm  250$&$ 1209\pm   42$&$  755\pm  138$&$   4.8^{+  1.0}_{-  0.1}$&$   1.2^{+  0.3}_{-  0.1}$&$   2.1^{+  0.8}_{-  0.7}$&$   0.5^{+  0.2}_{-  0.2}$&1,2 \\
NGC 3227&$\cdots$& 0.46&$ 92\pm 6$&PB&$ 42.36\pm 0.03$&$ 10.6^{+  6.1}_{-  6.1}$&$ 4445\pm  134$&$ 5278\pm 1117$&$ 1914\pm   71$&$ 2018\pm  174$&$  40.9^{+ 23.7}_{- 23.7}$&$   7.6^{+  4.4}_{-  4.4}$&$  57.6^{+ 41.1}_{- 41.1}$&$   8.4^{+  5.0}_{-  5.0}$&1,2 \\
NGC 3516&$\cdots$& 0.66&$181\pm 5$&PB&$ 43.17\pm 0.15$&$ 14.6^{+  1.4}_{-  1.1}$&  $\cdots$     &$ 5175\pm   96$&  $\cdots$     &$ 1591\pm   19$&$   \cdots $&$   \cdots$&$  76.3^{+  7.8}_{-  6.4}$&$   7.2^{+  0.7}_{-  0.6}$&1,2 \\
             &&     &          &  &$ 42.79\pm 0.20^c$&$ 11.7^{+  1.0d}_{-  1.5}$&$ 5384\pm  269^c$&  $\cdots$     &$ 2201\pm  110^c$&  $\cdots$     &$  66.2^{+  8.7}_{- 10.8}$&$  11.1^{+  1.5}_{-  1.8}$&$   \cdots$&$   \cdots$&2,5 \\
NGC 3783&$\cdots$& 0.04&$ 95\pm10$&PB&$ 42.55\pm 0.18$&$  7.3^{+  0.3}_{-  0.7}$&$ 3770\pm   68$&$ 3093\pm  529$&$ 1691\pm   19$&$ 1753\pm  141$&$  20.2^{+  1.1}_{-  2.1}$&$   4.1^{+  0.2}_{-  0.4}$&$  13.6^{+  4.7}_{-  4.8}$&$   4.4^{+  0.7}_{-  0.8}$&1,2 \\
NGC 4051&$\cdots$& 1.18&$ 89\pm 3$&PB&$ 41.96\pm 0.29$&$  2.5^{+  0.1}_{-  0.1}$&  $\cdots$     &$ 1034\pm   41$&  $\cdots$     &$  927\pm   64$&$   \cdots$&$   \cdots$&$   0.5^{+  0.04}_{-  0.04}$&$   0.4^{+  0.06}_{-  0.06}$&1,2 \\
             &&     &          &  &$ 41.96\pm 0.19^c$&$  1.9^{+  0.5d}_{-  0.5}$&$ 1072\pm  112^c$&  $\cdots$     &$  543\pm   52^c$&  $\cdots$     &$   0.4^{+  0.1}_{-  0.1}$&$   0.1^{+  0.03}_{-  0.03}$&$\cdots$&$ \cdots$&2,5 \\
NGC 4151&$\cdots$& 0.22&$ 97\pm 3$&CB&$ 42.09\pm 0.22$&$  6.0^{+  0.6}_{-  0.2}$&  $\cdots$     &$ 4711\pm  750$&  $\cdots$     &$ 2680\pm   64$&$   \cdots$&$   \cdots$&$  26.0^{+  8.7}_{-  8.3}$&$   8.4^{+  0.9}_{-  0.5}$&1,2 \\
             &&     &          &  &$ 43.05\pm 0.03^e$&$  3.1^{+  1.3e}_{-  1.3}$&$ 6371\pm  150^e$&  $\cdots$     &$ 2311\pm   11^e$&  $\cdots$     &$  24.6^{+ 10.4}_{- 10.4}$&$   3.2^{+  1.3}_{-  1.3}$&$\cdots$&$   \cdots$&2,6 \\
NGC 4253&Mrk 766& 0.99&$ 93\pm32$&PB&$ 42.57\pm 0.13$&$  5.4^{+  0.2}_{-  0.8}$&$ 1609\pm   39$&$  834\pm 1260$&$ 1088\pm   37$&$  516\pm  218$&$   2.7^{+  0.2}_{-  0.4}$&$   1.2^{+  0.1}_{-  0.2}$&$   0.7^{+  2.1}_{-  2.1}$&$   0.3^{+  0.3}_{-  0.3}$&1,2 \\
NGC 4593&Mrk 1330& 0.89&$135\pm 6$&PB&$ 42.79\pm 0.18$&$  4.5^{+  0.7}_{-  0.6}$&$ 5143\pm   16$&$ 4141\pm  416$&$ 1790\pm    3$&$ 1561\pm   55$&$  23.2^{+  3.6}_{-  3.1}$&$   2.8^{+  0.4}_{-  0.4}$&$  15.1^{+  3.8}_{-  3.6}$&$   2.1^{+  0.4}_{-  0.3}$&1,2 \\
NGC 4748&$\cdots$& 0.99&$105\pm13$&PB&$ 42.55\pm 0.13$&$  8.6^{+  0.6}_{-  0.4}$&$ 1947\pm   66$&$ 1212\pm  173$&$ 1009\pm   27$&$  657\pm   91$&$   6.4^{+  0.6}_{-  0.5}$&$   1.7^{+  0.1}_{-  0.1}$&$   2.5^{+  0.7}_{-  0.7}$&$   0.7^{+  0.2}_{-  0.2}$&1,2 \\
NGC 5548&$\cdots$& 0.10&$195\pm13$&CB&$ 42.95\pm 0.11$&$  5.5^{+  0.6}_{-  0.7}$&$12771\pm   71$&$11177\pm 2266$&$ 4266\pm   65$&$ 4270\pm  292$&$ 175.1^{+ 19.2}_{- 22.4}$&$  19.5^{+  2.2}_{-  2.6}$&$ 134.1^{+ 56.3}_{- 57.0}$&$  19.6^{+  3.4}_{-  3.7}$&1,2 \\
NGC 6814&$\cdots$& 0.45&$ 95\pm 3$&PB&$ 42.08\pm 0.29$&$  7.4^{+  0.1}_{-  0.1}$&$ 3323\pm    7$&$ 3277\pm  297$&$ 1918\pm   36$&$ 1610\pm  108$&$  15.9^{+  0.2}_{-  0.2}$&$   5.3^{+  0.2}_{-  0.2}$&$  15.5^{+  2.8}_{-  2.8}$&$   3.7^{+  0.5}_{-  0.5}$&1,2 \\
NGC 7469&$\cdots$& 0.60&$131\pm 5$&PB&$ 43.36\pm 0.10$&$ 11.7^{+  0.5}_{-  0.7}$&$ 1722\pm   30$&$ 2169\pm  459$&$ 1707\pm   20$&$ 1456\pm  207$&$   6.8^{+  0.4}_{-  0.5}$&$   6.7^{+  0.3}_{-  0.4}$&$  10.7^{+  4.6}_{-  4.6}$&$   4.8^{+  1.4}_{-  1.4}$&1,2 \\
PG  0921+525&Mrk 110& 0.14&$ 91\pm 7$&E &$ 43.62\pm 0.04$&$ 24.4^{+  2.2}_{- 12.7}$&$ 1543\pm    5$&$ 1494\pm  802$&$  962\pm   15$&$ 1196\pm  141$&$  11.3^{+  1.0}_{-  5.9}$&$   4.4^{+  0.4}_{-  2.3}$&$  10.6^{+ 11.4}_{- 12.6}$&$   6.8^{+  1.7}_{-  3.9}$&1,2 \\
             &&     &          &  &$ 43.69\pm 0.04$&$ 32.7^{+  5.9}_{-  5.2}$&$ 1658\pm    3$&$ 1381\pm  528$&$  953\pm   10$&$ 1115\pm  103$&$  17.5^{+  3.2}_{-  2.8}$&$   5.8^{+  1.1}_{-  0.9}$&$  12.2^{+  9.6}_{-  9.5}$&$   7.9^{+  2.0}_{-  1.9}$&1,2 \\
             &&     &          &  &$ 43.47\pm 0.05$&$ 20.8^{+  2.1}_{-  2.0}$&$ 1600\pm   39$&$ 1521\pm   59$&$  987\pm   18$&$  755\pm   29$&$  10.4^{+  1.2}_{-  1.1}$&$   4.0^{+  0.4}_{-  0.4}$&$   9.4^{+  1.2}_{-  1.2}$&$   2.3^{+  0.3}_{-  0.3}$&1,2 \\
PG  1229+204&Mrk 771& 0.53&$162\pm32$&PB&$ 43.64\pm 0.06$&$ 42.8^{+  2.3}_{-  1.1}$&$ 3828\pm   54$&$ 3415\pm  320$&$ 1608\pm   24$&$ 1385\pm  111$&$ 122.4^{+  7.4}_{-  4.7}$&$  21.6^{+  1.3}_{-  0.9}$&$  97.4^{+ 19.0}_{- 18.4}$&$  16.0^{+  2.7}_{-  2.6}$&1,2 \\
PG  1351+695&Mrk 279& 0.55&$197\pm12$&PB&$ 43.64\pm 0.08$&$ 17.8^{+  1.2}_{-  1.1}$&$ 5354\pm   32$&$ 3385\pm  349$&$ 1823\pm   11$&$ 1420\pm   96$&$  99.6^{+  6.8}_{-  6.3}$&$  11.5^{+  0.8}_{-  0.7}$&$  39.8^{+  8.6}_{-  8.6}$&$   7.0^{+  1.1}_{-  1.0}$&1,2 \\
PG  1411+442&$\cdots$& 0.63&$209\pm30$&CB&$ 44.50\pm 0.02$&$ 53.5^{+ 13.1}_{-  5.3}$&$ 2801\pm   43$&$ 2398\pm  353$&$ 1774\pm   29$&$ 1607\pm  169$&$  81.9^{+ 20.2}_{-  8.5}$&$  32.9^{+  8.1}_{-  3.4}$&$  60.0^{+ 23.0}_{- 18.6}$&$  27.0^{+  8.7}_{-  6.3}$&1,2 \\
PG  1426+015&Mrk 1383& 0.46&$217\pm15$&E &$ 44.57\pm 0.02$&$161.6^{+  6.9}_{- 11.1}$&$ 7113\pm  160$&$ 6323\pm 1295$&$ 2906\pm   80$&$ 3442\pm  308$&$1595.6^{+ 99.0}_{-131.0}$&$ 266.3^{+ 18.6}_{- 23.4}$&$1260.9^{+519.3}_{-523.7}$&$ 373.6^{+ 68.7}_{- 71.6}$&1,2 \\
PG  1434+590&Mrk 817& 0.69&$120\pm15$&PB&$ 43.73\pm 0.05$&$ 20.3^{+  2.2}_{-  2.2}$&$ 4711\pm   49$&$ 3515\pm  393$&$ 1984\pm    8$&$ 1392\pm   78$&$  87.9^{+  9.7}_{-  9.7}$&$  15.6^{+  1.7}_{-  1.7}$&$  48.9^{+ 12.2}_{- 12.2}$&$   7.7^{+  1.2}_{-  1.2}$&1,2 \\
             &&     &          &  &$ 43.61\pm 0.05$&$ 16.7^{+  1.8}_{-  2.6}$&$ 5237\pm   67$&$ 4952\pm  537$&$ 2098\pm   13$&$ 1971\pm   96$&$  89.4^{+  9.9}_{- 14.1}$&$  14.3^{+  1.6}_{-  2.2}$&$  79.9^{+ 19.4}_{- 21.3}$&$  12.7^{+  1.8}_{-  2.3}$&1,2 \\
             &&     &          &  &$ 43.61\pm 0.05$&$ 34.8^{+  4.7}_{-  5.6}$&$ 4767\pm   72$&$ 3752\pm  995$&$ 2195\pm   16$&$ 1729\pm  158$&$ 154.3^{+ 21.4}_{- 25.3}$&$  32.7^{+  4.4}_{-  5.3}$&$  95.6^{+ 52.3}_{- 53.0}$&$  20.3^{+  4.6}_{-  4.9}$&1,2 \\
PG  1534+580&Mrk 290& 0.29&$110\pm 5$&CB&$ 43.00\pm 0.08$&$  7.7^{+  0.7}_{-  0.5}$&$ 4543\pm  227^c$&$ 4270\pm  157^c$&$ 1769\pm   88$&$ 1609\pm   47$&$  31.0^{+  4.2}_{-  3.7}$&$   4.7^{+  0.6}_{-  0.6}$&$  27.4^{+  3.2}_{-  2.7}$&$   3.9^{+  0.4}_{-  0.3}$&1,2 \\
PG  1617+175&Mrk 877& 0.74&$201\pm37$&E &$ 44.33\pm 0.02$&$ 88.2^{+ 31.0}_{-  5.9}$&$ 6641\pm  190$&$ 4718\pm  991$&$ 2313\pm   69$&$ 2626\pm  211$&$ 759.1^{+270.3}_{- 66.8}$&$  92.1^{+ 32.8}_{-  8.3}$&$ 383.2^{+209.9}_{-163.0}$&$ 118.7^{+ 45.9}_{- 20.7}$&1,2 \\
PG  2130+099&Mrk 1513& 0.96&$163\pm19$&PB&$ 44.14\pm 0.03$&$ 35.0^{+  5.0}_{-  5.0}$&$ 1781\pm    5$&$ 2097\pm  102$&$ 1760\pm    2$&$ 1825\pm   65$&$  21.7^{+  3.1}_{-  3.1}$&$  21.2^{+  3.0}_{-  3.0}$&$  30.0^{+  5.2}_{-  5.2}$&$  22.8^{+  3.6}_{-  3.6}$&1,2 \\
SBS 1116+583A&    $\cdots$ & 0.59&$ 92\pm 4$&PB&$ 42.07\pm 0.28$&$  2.4^{+  0.9}_{-  0.9}$&$ 3668\pm  186$&$ 3604\pm 1123$&$ 1552\pm   36$&$ 1528\pm  184$&$   6.3^{+  2.4}_{-  2.4}$&$   1.1^{+  0.4}_{-  0.4}$&$   6.1^{+  4.4}_{-  4.4}$&$   1.1^{+  0.5}_{-  0.5}$&1,2 \\
Fairall 9        &    $\cdots$ & 0.49&$228\pm20^i$&CB&$ 43.92\pm 0.05$&$ 19.4^{+ 42.1}_{-  3.9}$&$ 6000\pm   66$&$ 6901\pm  707$&$ 2347\pm   16$&$ 3787\pm  197$&$ 136.0^{+296.0}_{- 28.0}$&$  21.0^{+ 45.0}_{-  4.2}$&$ 180.0^{+393.0}_{- 52.0}$&$  54.0^{+118.0}_{- 12.0}$&1,2,10 \\
NGC 5273&$\cdots$& 0.58&$ 74\pm 4^f$&E$^f$ &$ 41.53\pm 0.14^f$&$  2.2^{+  1.2f}_{-  1.6}$&$ 5688\pm  163^f$&$ 4615\pm  330^f$&$ 1821\pm   53^f$&$ 1544\pm   98^f$&$  14.0^{+  7.5}_{- 10.2}$&$   1.4^{+  0.8}_{-  1.0}$&$   9.2^{+  5.1}_{-  6.8}$&$   1.0^{+  0.6}_{-  0.7}$&2,7 \\
MCG+06-26-01 &    $\cdots$ & 1.04&$112\pm15^g$& PB$^h$&$ 42.67\pm 0.11^d$&$ 24.0^{+  8.4d}_{-  4.8}$&$ 1334\pm   80^d$&  $\cdots$     &$  785\pm   21^d$&  $\cdots$     &$  8.3^{+  3.1}_{-  1.9}$&$  2.9^{+  1.0}_{-  0.6}$& $\cdots$   & $\cdots$   &2,5,8,9 \\

\hline
\end{tabular}
\label{tab1}
\end{lrbox}
\scalebox{0.6}{\usebox{\tablebox}}

a. The superscript a indicates the data comes from \cite{Williams2018} 4.2.
b. The superscript b indicates the data comes from \cite{Ba15} Table 5.
c. The superscript c indicates the data comes from \cite{Du16} Table 1.
d. The superscript d indicates the data comes from \cite{Du15} Table 6 and Table 7.
e. The superscript e indicates the data comes from \cite{Co06} Table 1.
f.  The superscript f indicates the data comes from \cite{Be14}.
g. The superscript g indicates the data comes from \cite{Wo15} Table 1.
h. The superscript h indicates the bulge type derived from \cite{Wang2014}.

i. The superscript i indicates the data comes from \cite{O95}.

References: (1) \cite{HK14}, (2) \cite{Du16}, (3) \cite{Williams2018}, (4) \cite{Ba15},
(5) \cite{Du15}, (6) \cite{Co06}, (7) \cite{Be14}, (8) \cite{Wo15}, (9) \cite{Wang2014}, (10) \cite{O95}
\end{table*}

\begin{table*}
\caption{The high-z RM AGNs observed by SDSS with measured \sst.  }
\begin{lrbox}{\tablebox}
\begin{tabular}{llllllllllll}
\hline
Line&RMID&z&\sst&\fwm&\fwr&$\sigma_{\rm mean}$&$\sigma_{\rm rms}$&\vpfm&\vpsm&\vpfr&\vpsr\\
&&&\kms&\kms&\kms&\kms&\kms&$10^6\msun$&$10^6\msun$&$10^6\msun$&$10^6\msun$\\
\hline
\hb&RM  17&0.456&$ 191.4\pm 3.7$&$ 16318\pm 30$&$  7758\pm 77$&$  6937\pm 14$&$  6101\pm 48$&$1325.1^{+566.4}_{-301.4}$&$ 239.5^{+102.4}_{- 54.5}$&$ 299.5^{+128.0}_{- 68.1}$&$ 185.2^{+ 79.2}_{- 42.1}$\\
\hb&RM  33&0.715&$ 182.4\pm21.7$&$  1070\pm 30$&$  1626\pm243$&$   776\pm 13$&$   857\pm 32$&$   5.9^{+  2.2}_{-  2.0}$&$   3.1^{+  1.2}_{-  1.0}$&$  13.7^{+  5.1}_{-  4.5}$&$   3.8^{+  1.4}_{-  1.3}$\\
\hb&RM 177&0.482&$ 171.5\pm10.7$&$  5277\pm 39$&$  4930\pm163$&$  2541\pm  9$&$  2036\pm 39$&$  54.9^{+ 67.9}_{- 14.7}$&$  12.7^{+ 15.7}_{-  3.4}$&$  47.9^{+ 59.3}_{- 12.8}$&$   8.2^{+ 10.1}_{-  2.2}$\\
\hb&RM 191&0.442&$ 152.0\pm 8.5$&$  1316\pm 94$&$  1967\pm 76$&$   845\pm 12$&$  1030\pm 18$&$   2.9^{+  0.9}_{-  0.5}$&$   1.2^{+  0.4}_{-  0.2}$&$   6.4^{+  1.9}_{-  1.1}$&$   1.8^{+  0.5}_{-  0.3}$\\
\hb&RM 229&0.47 &$ 130.2\pm 8.7$&$  3055\pm180$&$  2377\pm288$&$  1722\pm 18$&$  1781\pm 38$&$  29.5^{+  5.3}_{-  8.2}$&$   9.4^{+  1.7}_{-  2.6}$&$  17.9^{+  3.2}_{-  5.0}$&$  10.0^{+  1.8}_{-  2.8}$\\
\hb&RM 267&0.587&$  97.1\pm 9.0$&$  2647\pm 23$&$  1998\pm 75$&$  1305\pm  6$&$  1202\pm 33$&$  27.9^{+  3.4}_{-  2.7}$&$   6.8^{+  0.8}_{-  0.7}$&$  15.9^{+  1.9}_{-  1.6}$&$   5.8^{+  0.7}_{-  0.6}$\\
\hb&RM 300&0.646&$ 109.4\pm11.9$&$  2110\pm 36$&$  2553\pm136$&$  1153\pm  8$&$  1232\pm 30$&$  26.4^{+  3.4}_{-  7.2}$&$   7.9^{+  1.0}_{-  2.2}$&$  38.7^{+  5.0}_{- 10.6}$&$   9.0^{+  1.2}_{-  2.5}$\\
\hb&RM 301&0.548&$ 176.9\pm10.1$&$ 18920\pm 91$&$ 10477\pm114$&$  7061\pm 25$&$  6259\pm 23$&$ 894.2^{+398.2}_{-314.4}$&$ 124.5^{+ 55.4}_{- 43.8}$&$ 274.2^{+122.1}_{- 96.4}$&$  97.9^{+ 43.6}_{- 34.4}$\\
\hb&RM 305&0.527&$ 150.5\pm 7.7$&$  2616\pm 21$&$  3172\pm 85$&$  2331\pm  7$&$  2126\pm 35$&$  71.5^{+  5.6}_{-  5.3}$&$  56.7^{+  4.5}_{-  4.2}$&$ 105.1^{+  8.3}_{-  7.9}$&$  47.2^{+  3.7}_{-  3.5}$\\
\hb&RM 320&0.265&$  66.4\pm 4.6$&$  3917\pm 28$&$  2718\pm 80$&$  1569\pm  9$&$  1462\pm 26$&$  75.5^{+ 14.1}_{- 17.1}$&$  12.1^{+  2.3}_{-  2.7}$&$  36.3^{+  6.8}_{-  8.2}$&$  10.5^{+  2.0}_{-  2.4}$\\
\hb&RM 338&0.418&$  83.3\pm 8.3$&$  4701\pm610$&$  5136\pm226$&$  2670\pm 28$&$  2291\pm 33$&$  46.1^{+ 24.1}_{- 19.0}$&$  14.9^{+  7.8}_{-  6.1}$&$  55.1^{+ 28.8}_{- 22.7}$&$  11.0^{+  5.8}_{-  4.5}$\\
\hb&RM 377&0.337&$ 115.3\pm 4.6$&$  3555\pm 42$&$  5654\pm239$&$  1648\pm 16$&$  1789\pm 23$&$  14.6^{+  1.0}_{-  1.5}$&$   3.1^{+  0.2}_{-  0.3}$&$  36.8^{+  2.5}_{-  3.7}$&$   3.7^{+  0.3}_{-  0.4}$\\
\hb&RM 392&0.843&$  77.2\pm25.6$&$  3540\pm199$&$ 10839\pm153$&$  3120\pm 46$&$  3658\pm 56$&$  34.7^{+  9.0}_{-  7.3}$&$  27.0^{+  7.0}_{-  5.7}$&$ 325.6^{+ 84.8}_{- 68.8}$&$  37.1^{+  9.7}_{-  7.8}$\\
\hb&RM 399&0.608&$ 187.2\pm 7.8$&$  2675\pm 60$&$  2578\pm112$&$  1429\pm 23$&$  1619\pm 38$&$  50.0^{+  1.5}_{- 14.4}$&$  14.3^{+  0.4}_{-  4.1}$&$  46.4^{+  1.4}_{- 13.3}$&$  18.3^{+  0.6}_{-  5.3}$\\
\hb&RM 457&0.604&$ 110.0\pm18.4$&$  6404\pm424$&$  7451\pm221$&$  2988\pm 83$&$  2788\pm 48$&$ 124.9^{+ 25.6}_{- 40.8}$&$  27.2^{+  5.6}_{-  8.9}$&$ 169.0^{+ 34.7}_{- 55.2}$&$  23.7^{+  4.9}_{-  7.7}$\\
\hb&RM 601&0.658&$ 214.9\pm19.2$&$ 16168\pm354$&$ 12673\pm455$&$  6705\pm 58$&$  5284\pm 54$&$ 591.8^{+438.7}_{-234.7}$&$ 101.8^{+ 75.5}_{- 40.4}$&$ 363.6^{+269.6}_{-144.2}$&$  63.2^{+ 46.9}_{- 25.1}$\\
\hb&RM 622&0.572&$ 122.9\pm 9.2$&$  2565\pm 36$&$  3234\pm164$&$  1369\pm  6$&$  1423\pm 32$&$  63.0^{+ 14.2}_{-  2.6}$&$  18.0^{+  4.1}_{-  0.7}$&$ 100.2^{+ 22.7}_{-  4.1}$&$  19.4^{+  4.4}_{-  0.8}$\\
\hb&RM 634&0.65 &$ 119.4\pm20.9$&$  1154\pm 42$&$  3422\pm491$&$  1059\pm 25$&$  1527\pm 22$&$   4.6^{+  2.2}_{-  1.9}$&$   3.9^{+  1.9}_{-  1.6}$&$  40.2^{+ 19.6}_{- 16.9}$&$   8.0^{+  3.9}_{-  3.4}$\\
\hb&RM 772&0.249&$ 136.5\pm 3.1$&$  2439\pm 33$&$  2078\pm 35$&$  1065\pm 14$&$  1026\pm 14$&$   4.5^{+  1.0}_{-  1.0}$&$   0.9^{+  0.2}_{-  0.2}$&$   3.3^{+  0.8}_{-  0.8}$&$   0.8^{+  0.2}_{-  0.2}$\\
\hb&RM 775&0.172&$ 130.4\pm 2.6$&$  3072\pm 24$&$  5010\pm 61$&$  1578\pm  5$&$  1818\pm  8$&$  30.0^{+ 24.1}_{- 12.1}$&$   7.9^{+  6.3}_{-  3.2}$&$  79.8^{+ 64.1}_{- 32.3}$&$  10.5^{+  8.4}_{-  4.3}$\\
\hb&RM 776&0.116&$ 112.4\pm 1.9$&$  3700\pm 16$&$  3111\pm 36$&$  1501\pm  5$&$  1409\pm 11$&$  28.1^{+  2.7}_{-  5.9}$&$   4.6^{+  0.4}_{-  1.0}$&$  19.8^{+  1.9}_{-  4.1}$&$   4.1^{+  0.4}_{-  0.9}$\\
\hb&RM 779&0.152&$  57.1\pm 4.9$&$  2670\pm 17$&$  2709\pm 55$&$  1249\pm  4$&$  1205\pm  9$&$  16.4^{+  1.0}_{-  2.1}$&$   3.6^{+  0.2}_{-  0.5}$&$  16.9^{+  1.0}_{-  2.1}$&$   3.3^{+  0.2}_{-  0.4}$\\
\hb&RM 781&0.263&$ 104.7\pm 4.3$&$  2515\pm 26$&$  3340\pm 82$&$  1169\pm  5$&$  1089\pm 22$&$  92.8^{+  3.9}_{-  4.1}$&$  20.1^{+  0.9}_{-  0.9}$&$ 163.7^{+  7.0}_{-  7.2}$&$  17.4^{+  0.7}_{-  0.8}$\\
\hb&RM 782&0.362&$ 129.5\pm 6.7$&$  3070\pm 49$&$  2730\pm137$&$  1378\pm  6$&$  1353\pm 23$&$  36.8^{+  2.0}_{-  5.5}$&$   7.4^{+  0.4}_{-  1.1}$&$  29.1^{+  1.6}_{-  4.4}$&$   7.1^{+  0.4}_{-  1.1}$\\
\hb&RM 790&0.237&$ 204.4\pm 3.1$&$ 17112\pm 81$&$  9448\pm367$&$  6813\pm 13$&$  6318\pm 38$&$ 314.3^{+325.7}_{-120.0}$&$  49.8^{+ 51.6}_{- 19.0}$&$  95.8^{+ 99.3}_{- 36.6}$&$  42.8^{+ 44.4}_{- 16.3}$\\
\hb&RM 840&0.244&$ 164.3\pm 3.6$&$ 15735\pm 93$&$  6580\pm 48$&$  6596\pm 22$&$  4457\pm 60$&$ 241.6^{+ 72.5}_{- 67.6}$&$  42.5^{+ 12.8}_{- 11.9}$&$  42.2^{+ 12.7}_{- 11.8}$&$  19.4^{+  5.8}_{-  5.4}$\\
\ha&RM  17&0.456&$ 191.4\pm 3.7$&$  4159\pm 13$&$  5604\pm 31$&$  4509\pm 53$&$  4569\pm 51$&$ 191.1^{+ 24.6}_{- 51.0}$&$ 224.6^{+ 29.0}_{- 59.9}$&$ 346.9^{+ 44.7}_{- 92.5}$&$ 230.6^{+ 29.7}_{- 61.5}$\\
\ha&RM  88&0.516&$ 128.5\pm12.3$&$  4451\pm 32$&$ 10290\pm142$&$  2449\pm 27$&$  3320\pm 26$&$ 211.9^{+ 11.2}_{- 19.7}$&$  64.1^{+  3.4}_{-  6.0}$&$1132.4^{+ 59.9}_{-105.4}$&$ 117.9^{+  6.2}_{- 11.0}$\\
\ha&RM 191&0.442&$ 152.0\pm 8.5$&$  2050\pm 18$&$  1575\pm 60$&$   858\pm  6$&$   796\pm 23$&$  13.7^{+  3.4}_{-  4.5}$&$   2.4^{+  0.6}_{-  0.8}$&$   8.1^{+  2.0}_{-  2.7}$&$   2.1^{+  0.5}_{-  0.7}$\\
\ha&RM 229&0.47 &$ 130.2\pm 8.7$&$  2271\pm 34$&$  2103\pm365$&$  1528\pm 10$&$  1738\pm 31$&$  22.2^{+  7.7}_{-  7.3}$&$  10.1^{+  3.5}_{-  3.3}$&$  19.1^{+  6.7}_{-  6.3}$&$  13.0^{+  4.5}_{-  4.3}$\\
\ha&RM 252&0.281&$ 180.7\pm 4.0$&$  6574\pm 69$&$  7868\pm 66$&$  4300\pm 26$&$  3384\pm 71$&$  85.2^{+ 20.2}_{- 16.0}$&$  36.4^{+  8.6}_{-  6.8}$&$ 122.0^{+ 29.0}_{- 23.0}$&$  22.6^{+  5.4}_{-  4.3}$\\
\ha&RM 320&0.265&$  66.4\pm 4.6$&$  3232\pm 12$&$  2808\pm 41$&$  1538\pm  3$&$  1320\pm 17$&$  41.2^{+ 21.4}_{- 19.0}$&$   9.3^{+  4.8}_{-  4.3}$&$  31.1^{+ 16.2}_{- 14.3}$&$   6.9^{+  3.6}_{-  3.2}$\\
\ha&RM 377&0.337&$ 115.3\pm 4.6$&$  2802\pm 17$&$  2971\pm114$&$  1407\pm 10$&$  1372\pm 40$&$   8.7^{+  0.8}_{-  0.8}$&$   2.2^{+  0.2}_{-  0.2}$&$   9.8^{+  0.9}_{-  0.9}$&$   2.1^{+  0.2}_{-  0.2}$\\
\ha&RM 733&0.455&$ 196.9\pm16.6$&$  3284\pm 21$&$  3489\pm 84$&$  1488\pm  7$&$  1590\pm 24$&$ 111.6^{+ 18.3}_{- 12.0}$&$  22.9^{+  3.8}_{-  2.5}$&$ 125.9^{+ 20.7}_{- 13.5}$&$  26.1^{+  4.3}_{-  2.8}$\\
\ha&RM 768&0.258&$ 171.9\pm 2.8$&$  6213\pm  9$&$  6279\pm 20$&$  3428\pm 16$&$  3232\pm 40$&$ 317.2^{+ 20.3}_{- 15.8}$&$  96.6^{+  6.2}_{-  4.8}$&$ 323.9^{+ 20.8}_{- 16.2}$&$  85.8^{+  5.5}_{-  4.3}$\\
\ha&RM 772&0.249&$ 136.5\pm 3.1$&$  2483\pm  9$&$  2142\pm 11$&$  1104\pm  2$&$   907\pm  6$&$   7.1^{+  1.9}_{-  1.2}$&$   1.4^{+  0.4}_{-  0.2}$&$   5.3^{+  1.4}_{-  0.9}$&$   0.9^{+  0.2}_{-  0.2}$\\
\ha&RM 776&0.116&$ 112.4\pm 1.9$&$  2877\pm  6$&$  2794\pm 15$&$  1426\pm  2$&$  1185\pm  7$&$  13.4^{+  7.9}_{-  3.7}$&$   3.3^{+  1.9}_{-  0.9}$&$  12.6^{+  7.4}_{-  3.5}$&$   2.3^{+  1.4}_{-  0.6}$\\
\ha&RM 779&0.152&$  57.1\pm 4.9$&$  2453\pm  5$&$  2643\pm 23$&$  1126\pm  2$&$  1018\pm  7$&$  94.2^{+  5.8}_{-  7.4}$&$  19.8^{+  1.2}_{-  1.6}$&$ 109.3^{+  6.7}_{-  8.6}$&$  16.2^{+  1.0}_{-  1.3}$\\
\ha&RM 790&0.237&$ 204.4\pm 3.1$&$  5769\pm 18$&$  8898\pm 66$&$  3532\pm 17$&$  5157\pm 40$&$ 292.3^{+153.9}_{- 25.3}$&$ 109.6^{+ 57.7}_{-  9.5}$&$ 695.3^{+366.2}_{- 60.3}$&$ 233.6^{+123.0}_{- 20.2}$\\
\ha&RM 840&0.244&$ 164.3\pm 3.6$&$  4593\pm 14$&$  6027\pm 19$&$  3002\pm 45$&$  3927\pm 30$&$  43.6^{+  9.5}_{-  9.9}$&$  18.6^{+  4.0}_{-  4.2}$&$  75.1^{+ 16.3}_{- 17.0}$&$  31.9^{+  6.9}_{-  7.2}$\\
\hline
\end{tabular}
\label{tab2}
\end{lrbox}
\scalebox{0.8}{\usebox{\tablebox}}
\end{table*}

\begin{table*}
\caption{The \mbh and four kinds factor of the low-z RM AGNs sample in Table \ref{tab1}.  }
\begin{lrbox}{\tablebox}
\begin{tabular}{lllllll}
\hline
Name &Type&$\log \mbh/\msun$&$\log f_{\rm F,mean} $&$\log f_{\rm \sigma, mean}$&$ \log f_{\rm F, rms} $&$ \log f_{\rm \sigma, rms}$\\
\hline
3C      120      &E &$    8.09\pm  0.23$           &$ 1.05^{+ 0.23}_{- 0.23}$                         &$ 0.91^{+ 0.23}_{- 0.23}$                         &$ 0.56^{+ 0.28}_{- 0.28}$                         &$ 1.00^{+ 0.24}_{- 0.24}$                         \\
3C      390.3    &E &$    9.08\pm  0.11$           &$-0.13^{+ 0.12}_{- 0.11}$                         &$ 0.65^{+ 0.12}_{- 0.11}$                         &$ 0.04^{+ 0.18}_{- 0.17}$                         &$ 0.64^{+ 0.12}_{- 0.12}$                         \\
Ark     120      &CB&$    8.41\pm  0.08$                         &$ 0.02^{+ 0.14}_{- 0.11}$                         &$ 1.09^{+ 0.13}_{- 0.11}$                         &$ 0.10^{+ 0.14}_{- 0.12}$                         &$ 1.00^{+ 0.14}_{- 0.12}$                         \\
                 &  &                &$ 0.07^{+ 0.12}_{- 0.09}$                         &$ 1.12^{+ 0.12}_{- 0.09}$                         &$ 0.22^{+ 0.12}_{- 0.10}$                         &$ 1.11^{+ 0.12}_{- 0.09}$                         \\
Arp     151      &CB&$    7.49\pm  0.06$           &$ 0.66^{+ 0.07}_{- 0.11}$                         &$ 1.04^{+ 0.07}_{- 0.10}$                         &$ 0.90^{+ 0.08}_{- 0.12}$                         &$ 1.44^{+ 0.07}_{- 0.10}$                         \\
Mrk     50       &CB&$    7.34\pm  0.24$           &$-0.12^{+ 0.25}_{- 0.26}$                         &$ 0.50^{+ 0.25}_{- 0.26}$                         &$ 0.06^{+ 0.26}_{- 0.26}$                         &$ 0.50^{+ 0.26}_{- 0.26}$                         \\
Mrk     79       &CB&$    7.67\pm  0.17$           &$-0.43^{+ 0.30}_{- 0.18}$                         &$ 0.25^{+ 0.30}_{- 0.18}$                         &$-0.43^{+ 0.38}_{- 0.30}$                         &$ 0.32^{+ 0.33}_{- 0.24}$                         \\
                 &  &                              &$-0.45^{+ 0.18}_{- 0.18}$                         &$ 0.18^{+ 0.18}_{- 0.18}$                         &$-0.35^{+ 0.18}_{- 0.18}$                         &$ 0.45^{+ 0.18}_{- 0.18}$                         \\
                 &  &                              &$-0.20^{+ 0.18}_{- 0.25}$                         &$ 0.43^{+ 0.18}_{- 0.25}$                         &$-0.29^{+ 0.20}_{- 0.27}$                         &$ 0.62^{+ 0.19}_{- 0.25}$                         \\
Mrk     202      &PB&$ 6.70\pm 0.07( 6.12\pm 0.07)$&$ 0.52^{+ 0.08}_{- 0.08}(-0.06^{+ 0.08}_{- 0.08})$&$ 1.00^{+ 0.07}_{- 0.07}( 0.42^{+ 0.07}_{- 0.07})$&$ 0.58^{+ 0.18}_{- 0.18}( 0.00^{+ 0.18}_{- 0.18})$&$ 1.22^{+ 0.16}_{- 0.16}( 0.64^{+ 0.16}_{- 0.16})$\\
Mrk     509      &E &$    8.33\pm  0.12$           &$ 0.25^{+ 0.12}_{- 0.12}$                         &$ 0.83^{+ 0.12}_{- 0.12}$                         &$ 0.34^{+ 0.13}_{- 0.13}$                         &$ 1.00^{+ 0.12}_{- 0.12}$                         \\
Mrk     590      &PB&$ 8.38\pm 0.06( 7.80\pm 0.06)$&$ 0.67^{+ 0.14}_{- 0.12}( 0.09^{+ 0.14}_{- 0.12})$&$ 1.14^{+ 0.11}_{- 0.07}( 0.56^{+ 0.11}_{- 0.07})$&$ 0.99^{+ 0.11}_{- 0.08}( 0.41^{+ 0.11}_{- 0.08})$&$ 1.24^{+ 0.11}_{- 0.08}( 0.66^{+ 0.11}_{- 0.08})$\\
                 &  &                              &$ 0.71^{+ 0.13}_{- 0.08}( 0.13^{+ 0.13}_{- 0.08})$&$ 1.00^{+ 0.13}_{- 0.08}( 0.42^{+ 0.13}_{- 0.08})$&$ 0.94^{+ 0.27}_{- 0.25}( 0.36^{+ 0.27}_{- 0.25})$&$ 1.40^{+ 0.14}_{- 0.09}( 0.82^{+ 0.14}_{- 0.09})$\\
                 &  &                              &$ 0.82^{+ 0.07}_{- 0.07}( 0.24^{+ 0.07}_{- 0.07})$&$ 1.06^{+ 0.07}_{- 0.07}( 0.48^{+ 0.07}_{- 0.07})$&$ 1.02^{+ 0.18}_{- 0.18}( 0.44^{+ 0.18}_{- 0.18})$&$ 1.45^{+ 0.12}_{- 0.12}( 0.87^{+ 0.12}_{- 0.12})$\\
Mrk     1310     &CB&$    6.84\pm  0.11$           &$ 0.16^{+ 0.11}_{- 0.14}$                         &$ 0.76^{+ 0.12}_{- 0.16}$                         &$ 0.52^{+ 0.18}_{- 0.20}$                         &$ 1.14^{+ 0.21}_{- 0.21}$                         \\
NGC     3227     &PB&$ 7.32\pm 0.18$&$-0.29^{+ 0.31}_{- 0.31}$&$ 0.44^{+ 0.31}_{- 0.31}$&$-0.44^{+ 0.36}_{- 0.36}$&$ 0.40^{+ 0.32}_{- 0.32}$\\
NGC     3516     &PB&$ 8.30\pm 0.05( 7.72\pm 0.05)$&$\cdots$                                          &$\cdots$                                          &$ 0.42^{+ 0.06}_{- 0.07}(-0.16^{+ 0.06}_{- 0.07})$&$ 1.44^{+ 0.06}_{- 0.07}( 0.86^{+ 0.06}_{- 0.07})$\\
                 &  &                              &$ 0.48^{+ 0.09}_{- 0.08}(-0.10^{+ 0.09}_{- 0.08})$&$ 1.25^{+ 0.09}_{- 0.08}( 0.67^{+ 0.09}_{- 0.08})$&$\cdots$                                          &$\cdots$                                          \\
NGC     3783     &PB&$ 7.07\pm 0.20( 6.49\pm 0.20)$&$-0.23^{+ 0.20}_{- 0.20}(-0.81^{+ 0.20}_{- 0.20})$&$ 0.46^{+ 0.20}_{- 0.20}(-0.12^{+ 0.20}_{- 0.20})$&$-0.06^{+ 0.25}_{- 0.25}(-0.64^{+ 0.25}_{- 0.25})$&$ 0.43^{+ 0.21}_{- 0.21}(-0.15^{+ 0.21}_{- 0.21})$\\
NGC     4051     &PB&$ 6.95\pm 0.06( 6.37\pm 0.06)$&$\cdots$                                          &$\cdots$                                          &$ 1.25^{+ 0.06}_{- 0.06}( 0.67^{+ 0.06}_{- 0.06})$&$ 1.35^{+ 0.12}_{- 0.12}( 0.77^{+ 0.12}_{- 0.12})$\\
                 &  &                              &$ 1.35^{+ 0.12}_{- 0.12}( 0.77^{+ 0.12}_{- 0.12})$&$ 1.95^{+ 0.06}_{- 0.06}( 1.37^{+ 0.06}_{- 0.06})$&$\cdots$                                          &$\cdots$                                          \\
NGC     4151     &CB&$    7.49\pm  0.06$           &$\cdots$                                          &$\cdots$                                          &$ 0.07^{+ 0.15}_{- 0.15}$                         &$ 0.56^{+ 0.06}_{- 0.07}$                         \\
                 &  &                              &$ 0.09^{+ 0.19}_{- 0.19}$                         &$ 0.98^{+ 0.18}_{- 0.18}$                         &$\cdots$                                          &$\cdots$                                          \\
NGC     4253     &PB&$ 7.03\pm 0.65( 6.45\pm 0.65)$&$ 0.60^{+ 0.65}_{- 0.65}( 0.02^{+ 0.65}_{- 0.65})$&$ 0.95^{+ 0.65}_{- 0.65}( 0.37^{+ 0.65}_{- 0.65})$&$ 1.19^{+ 1.44}_{- 1.44}( 0.61^{+ 1.44}_{- 1.44})$&$ 1.56^{+ 0.78}_{- 0.78}( 0.98^{+ 0.78}_{- 0.78})$\\
NGC     4593     &PB&$ 7.74\pm 0.08( 7.16\pm 0.08)$&$ 0.38^{+ 0.10}_{- 0.11}(-0.20^{+ 0.10}_{- 0.11})$&$ 1.30^{+ 0.10}_{- 0.10}( 0.72^{+ 0.10}_{- 0.10})$&$ 0.56^{+ 0.13}_{- 0.14}(-0.02^{+ 0.13}_{- 0.14})$&$ 1.42^{+ 0.10}_{- 0.12}( 0.84^{+ 0.10}_{- 0.12})$\\
NGC     4748     &PB&$ 7.26\pm 0.23( 6.68\pm 0.23)$&$ 0.46^{+ 0.24}_{- 0.24}(-0.12^{+ 0.24}_{- 0.24})$&$ 1.03^{+ 0.23}_{- 0.23}( 0.45^{+ 0.23}_{- 0.23})$&$ 0.87^{+ 0.26}_{- 0.26}( 0.29^{+ 0.26}_{- 0.26})$&$ 1.42^{+ 0.26}_{- 0.26}( 0.84^{+ 0.26}_{- 0.26})$\\
NGC     5548     &CB&$    8.44\pm  0.13$           &$ 0.20^{+ 0.14}_{- 0.13}$                         &$ 1.15^{+ 0.14}_{- 0.13}$                         &$ 0.31^{+ 0.22}_{- 0.22}$                         &$ 1.15^{+ 0.15}_{- 0.15}$                         \\
NGC     6814     &PB&$ 7.07\pm 0.06( 6.49\pm 0.06)$&$-0.13^{+ 0.06}_{- 0.06}(-0.71^{+ 0.06}_{- 0.06})$&$ 0.35^{+ 0.06}_{- 0.06}(-0.23^{+ 0.06}_{- 0.06})$&$-0.12^{+ 0.10}_{- 0.10}(-0.70^{+ 0.10}_{- 0.10})$&$ 0.51^{+ 0.08}_{- 0.08}(-0.07^{+ 0.08}_{- 0.08})$\\
NGC     7469     &PB&$ 7.69\pm 0.07( 7.11\pm 0.07)$&$ 0.85^{+ 0.08}_{- 0.08}( 0.27^{+ 0.08}_{- 0.08})$&$ 0.86^{+ 0.08}_{- 0.07}( 0.28^{+ 0.08}_{- 0.07})$&$ 0.66^{+ 0.20}_{- 0.20}( 0.08^{+ 0.20}_{- 0.20})$&$ 1.00^{+ 0.14}_{- 0.14}( 0.42^{+ 0.14}_{- 0.14})$\\
PG      0921+525 &E &$    6.99\pm  0.14$           &$-0.06^{+ 0.27}_{- 0.15}$                         &$ 0.35^{+ 0.27}_{- 0.15}$                         &$-0.03^{+ 0.53}_{- 0.48}$                         &$ 0.16^{+ 0.29}_{- 0.18}$                         \\
                 &  &                              &$-0.25^{+ 0.16}_{- 0.16}$                         &$ 0.23^{+ 0.16}_{- 0.17}$                         &$-0.09^{+ 0.36}_{- 0.37}$                         &$ 0.09^{+ 0.18}_{- 0.18}$                         \\
                 &  &                              &$-0.02^{+ 0.15}_{- 0.15}$                         &$ 0.39^{+ 0.15}_{- 0.15}$                         &$ 0.02^{+ 0.15}_{- 0.15}$                         &$ 0.63^{+ 0.16}_{- 0.16}$                         \\
PG      1229+204 &PB&$ 8.09\pm 0.37( 7.51\pm 0.37)$&$ 0.00^{+ 0.37}_{- 0.37}(-0.58^{+ 0.37}_{- 0.37})$&$ 0.75^{+ 0.37}_{- 0.37}( 0.17^{+ 0.37}_{- 0.37})$&$ 0.10^{+ 0.38}_{- 0.38}(-0.48^{+ 0.38}_{- 0.38})$&$ 0.89^{+ 0.38}_{- 0.38}( 0.31^{+ 0.38}_{- 0.38})$\\
PG      1351+695 &PB&$ 8.46\pm 0.11( 7.88\pm 0.11)$&$ 0.46^{+ 0.12}_{- 0.12}(-0.12^{+ 0.12}_{- 0.12})$&$ 1.40^{+ 0.12}_{- 0.12}( 0.82^{+ 0.12}_{- 0.12})$&$ 0.86^{+ 0.15}_{- 0.15}( 0.28^{+ 0.15}_{- 0.15})$&$ 1.62^{+ 0.13}_{- 0.13}( 1.04^{+ 0.13}_{- 0.13})$\\
PG      1411+442 &CB&$    8.57\pm  0.27$           &$ 0.66^{+ 0.27}_{- 0.29}$                         &$ 1.06^{+ 0.27}_{- 0.29}$                         &$ 0.80^{+ 0.30}_{- 0.32}$                         &$ 1.14^{+ 0.29}_{- 0.30}$                         \\
PG      1426+015 &E &$    8.65\pm  0.13$           &$-0.56^{+ 0.13}_{- 0.13}$                         &$ 0.22^{+ 0.14}_{- 0.13}$                         &$-0.46^{+ 0.22}_{- 0.22}$                         &$ 0.07^{+ 0.15}_{- 0.15}$                         \\
PG      1434+590 &PB&$ 7.52\pm 0.24( 6.94\pm 0.24)$&$-0.43^{+ 0.24}_{- 0.24}(-1.01^{+ 0.24}_{- 0.24})$&$ 0.33^{+ 0.24}_{- 0.24}(-0.25^{+ 0.24}_{- 0.24})$&$-0.17^{+ 0.26}_{- 0.26}(-0.75^{+ 0.26}_{- 0.26})$&$ 0.63^{+ 0.24}_{- 0.24}( 0.05^{+ 0.24}_{- 0.24})$\\
                 &  &                              &$-0.43^{+ 0.24}_{- 0.24}(-1.01^{+ 0.24}_{- 0.24})$&$ 0.36^{+ 0.24}_{- 0.24}(-0.22^{+ 0.24}_{- 0.24})$&$-0.38^{+ 0.26}_{- 0.26}(-0.96^{+ 0.26}_{- 0.26})$&$ 0.41^{+ 0.25}_{- 0.24}(-0.17^{+ 0.25}_{- 0.24})$\\
                 &  &                              &$-0.67^{+ 0.25}_{- 0.24}(-1.25^{+ 0.25}_{- 0.24})$&$ 0.00^{+ 0.25}_{- 0.24}(-0.58^{+ 0.25}_{- 0.24})$&$-0.46^{+ 0.34}_{- 0.33}(-1.04^{+ 0.34}_{- 0.33})$&$ 0.21^{+ 0.26}_{- 0.25}(-0.37^{+ 0.26}_{- 0.25})$\\
PG      1534+580 &CB&$    7.35\pm  0.09$           &$-0.14^{+ 0.10}_{- 0.10}$                         &$ 0.68^{+ 0.10}_{- 0.10}$                         &$-0.09^{+ 0.10}_{- 0.10}$                         &$ 0.76^{+ 0.09}_{- 0.10}$                         \\
PG      1617+175 &E &$    8.50\pm  0.35$           &$-0.38^{+ 0.35}_{- 0.38}$                         &$ 0.54^{+ 0.35}_{- 0.38}$                         &$-0.08^{+ 0.39}_{- 0.42}$                         &$ 0.42^{+ 0.35}_{- 0.38}$                         \\
PG      2130+099 &PB&$ 8.10\pm 0.22( 7.52\pm 0.22)$&$ 0.76^{+ 0.23}_{- 0.23}( 0.18^{+ 0.23}_{- 0.23})$&$ 0.77^{+ 0.23}_{- 0.23}( 0.19^{+ 0.23}_{- 0.23})$&$ 0.62^{+ 0.23}_{- 0.23}( 0.04^{+ 0.23}_{- 0.23})$&$ 0.74^{+ 0.23}_{- 0.23}( 0.16^{+ 0.23}_{- 0.23})$\\
SBS     1116+583A&PB&$ 7.01\pm 0.08( 6.43\pm 0.08)$&$ 0.21^{+ 0.18}_{- 0.18}(-0.37^{+ 0.18}_{- 0.18})$&$ 0.97^{+ 0.18}_{- 0.18}( 0.39^{+ 0.18}_{- 0.18})$&$ 0.23^{+ 0.32}_{- 0.32}(-0.35^{+ 0.32}_{- 0.32})$&$ 0.97^{+ 0.21}_{- 0.21}( 0.39^{+ 0.21}_{- 0.21})$\\
Fairall 9        &CB&$    8.74\pm  0.17$           &$ 0.61^{+ 0.19}_{- 0.95}$                         &$ 1.42^{+ 0.19}_{- 0.94}$                         &$ 0.48^{+ 0.21}_{- 0.95}$                         &$ 1.01^{+ 0.19}_{- 0.95}$                         \\
NGC     5273     &E &$    6.60\pm  0.09$           &$-0.54^{+ 0.33}_{- 0.25}$                         &$ 0.46^{+ 0.32}_{- 0.26}$                         &$-0.36^{+ 0.33}_{- 0.26}$                         &$ 0.60^{+ 0.32}_{- 0.27}$                         \\
MCG     +06-26-01&PB&$ 7.39\pm 0.25( 6.81\pm 0.25)$&$ 0.47^{+ 0.27}_{- 0.30}(-0.11^{+ 0.27}_{- 0.30})$&$ 0.92^{+ 0.27}_{- 0.29}( 0.34^{+ 0.27}_{- 0.29})$&$\cdots$                                          &$\cdots$                                          \\

\hline
\end{tabular}
\label{tab3}
\end{lrbox}
\scalebox{0.8}{\usebox{\tablebox}}
\\a. The superscript a indicates the mass of SMBH is measured by dynamic method comes from \cite{Da06}.
b. The superscript b indicates the mass of SMBH is measured by dynamic method comes from \cite{On14} Table 1.
\mbh and $f$ are  derived from two kinds of \ms relation (Equation \ref{eq1}, $\beta = 4.38$ and $\alpha = -0.51$ or $\alpha=-1.09$ (in brackets) )
\end{table*}

\begin{table*}
\begin{center}
\caption{Distributions of the virial coefficient for different populations for low-z AGN RM sample in Table \ref{tab1}.
Pop1: $\dhb(mean) < 2.35$, Pop2: \dhb(mean) $\ge$ 2.35. PopA: $\fwm < 4000\kms$,
PopB: $\fwm \ge 4000 \kms$. For lines of K-S test, the values are d, $prob(\rm KS)$. A small value of $prob(\rm KS)$ shows a significantly different between two sets of data. Values of $prob(\rm KS) < 0.01$ are highlighted in boldface.}
\begin{tabular}{lcccccccc}
\hline
 & \multicolumn{2}{c}{Ellipticals and Classical bulges} & n & \multicolumn{2}{c}{ALL(PB unscaled)} & \multicolumn{2}{c}{ALL(PB scaled by 1/3.80)} & n \\
\hline
\multicolumn{9}{c}{MEAN SPECTRUM} \\ \hline
& $\log f_{\rm \sigma,mean}$ & $\log f_{\rm F,mean}$ &  & $\log f_{\rm \sigma,mean}$ &  $\log f_{\rm F,mean}$ & $\log f_{\rm \sigma,mean}$ & $\log f_{\rm F,mean}$ &  \\
\hline  
total & $ 0.76\pm 0.34$&$ 0.08\pm 0.46$& 17 &$ 0.84\pm 0.39$&$ 0.22\pm 0.48$& $ 0.57\pm 0.42$&$-0.05\pm 0.46$ & 34 \\
Pop1 & $ 0.71\pm 0.31$&$ 0.27\pm 0.48$& 8 & $ 0.79\pm 0.40$&$ 0.34\pm 0.52$ & $ 0.47\pm 0.43$&$ 0.02\pm 0.53$ & 20 \\
Pop2 & $ 0.80\pm 0.39$&$-0.09\pm 0.38$& 9 & $ 0.92\pm 0.38$&$ 0.05\pm 0.37$&  $ 0.71\pm 0.36$&$-0.16\pm 0.33$ & 14 \\

PopA & $ 0.82\pm 0.27$&$ 0.45\pm 0.42$ & 6 & $ 0.89\pm 0.35$&$ 0.46\pm 0.43$ & $ 0.50\pm 0.42$&$ 0.07\pm 0.51$ & 18\\
PopB & $ 0.73\pm 0.39$&$-0.12\pm 0.35$ & 11 & $ 0.79\pm 0.43$&$-0.05\pm 0.39$& $ 0.64\pm 0.42$&$-0.19\pm 0.37$& 16 \\ \hline
K-S Pop1-Pop2 & 0.33, 0.64&0.43, 0.31&&0.38, 0.14&0.38, 0.14 &	0.41, \bfseries 0.09&0.32, 0.30 \\
K-S PopA-PopB & 0.47, 0.26&0.65, \bfseries 0.04&&	0.40, 0.10&0.47, \bfseries 0.03	&0.37, 0.16&0.41, \bfseries 0.08\\			
\hline
\multicolumn{9}{c}{RMS SPECTRUM} \\ \hline
& $\log f_{\rm \sigma,rms}$ & $\log f_{\rm F,rms}$ &  & $\log f_{\rm \sigma,rms}$ & $\log f_{\rm F,rms}$ & $\log f_{\rm \sigma,rms}$ & $\log f_{\rm F,rms}$ &  \\
\hline 
total & $ 0.78\pm 0.37$&$ 0.17\pm 0.40$ & 17 & $ 0.91\pm 0.42$&$ 0.31\pm 0.48$&$ 0.65\pm 0.41$&$ 0.05\pm 0.45$ & 33 \\
Pop1 &  $ 0.87\pm 0.41$&$ 0.35\pm 0.43$ & 8 & $ 0.92\pm 0.43$&$ 0.42\pm 0.53$& $ 0.61\pm 0.46$&$ 0.11\pm 0.53$ & 19 \\
Pop2 & $ 0.70\pm 0.34$&$ 0.01\pm 0.30$ & 9 & $ 0.90\pm 0.43$&$ 0.16\pm 0.36$&$ 0.69\pm 0.32$&$-0.05\pm 0.30$ & 14 \\

PopA & $ 1.00\pm 0.38$&$ 0.51\pm 0.33$& 6 & $ 1.03\pm 0.37$&$ 0.55\pm 0.43$ & $ 0.65\pm 0.45$&$ 0.18\pm 0.50$ & 17 \\
PopB & $ 0.66\pm 0.32$&$-0.02\pm 0.29$ & 11 & $ 0.78\pm 0.45$&$ 0.05\pm 0.40$& $ 0.64\pm 0.36$&$-0.09\pm 0.35$& 16 \\ \hline
K-S Pop1-Pop2 &0.29, 0.79&0.51, 0.14&&0.23, 0.75&0.44, \bfseries 0.07 &0.19, 0.90&0.35, 0.22		\\
K-S PopA-PopB &0.56, 0.11&0.74, \bfseries 0.01&&0.39, 0.12&0.52, \bfseries 0.01 & 0.23, 0.73& 0.41, \bfseries 0.09		 \\

\hline
\end{tabular}
\label{tab4}
\end{center}
\end{table*}

\begin{table*}
\begin{center}
\caption{The Spearman correlation coefficient and the uncorrelated probability of virial coefficient versus the others parameters for low-z AGN RM sample in Table \ref{tab1}.
 $R_s$ is the Spearman correlation coefficient and $p_{\rm null}$ is the probability of the null hypothesis. Values of $R_s > 0.5 $ or $p_{\rm null} < 0.01$ are highlighted in boldface. }
\begin{tabular}{lcccccccccccc}
\hline
 & \multicolumn{4}{c}{Elliptical and Classical bulges} & \multicolumn{4}{c}{ALL(PB unscaled)} & \multicolumn{4}{c}{ALL(PB scaled by 1/3.80)}\\
\hline
& $R_{s}$ & $p_{\rm null}$ & $R_{s}$ & $p_{\rm null}$ & $R_{s}$ & $p_{\rm null}$ & $R_{s}$ & $p_{\rm null}$&$R_{s}$ & $p_{\rm null}$  & $R_{s}$ & $p_{\rm null}$ \\
\hline
\multicolumn{13}{c}{MEAN SPECTRUM} \\
\hline
 & \multicolumn{2}{c}{$f_{\rm \sigma, mean}$} & \multicolumn{2}{c}{$f_{\rm F. mean}$} & \multicolumn{2}{c}{$f_{\rm \sigma, mean}$} & \multicolumn{2}{c}{$f_{\rm F, mean}$}& \multicolumn{2}{c}{$f_{\rm \sigma, mean}$} & \multicolumn{2}{c}{$f_{\rm F, mean}$} \\
 \hline
\sst &0.29  & 2.60E-01 & 0.11  & 6.90E-01 & 0.23  & 1.90E-01 & 0.09  & 6.10E-01 & 0.28  & 1.10E-01 & 0.17  & 3.40E-01 \\
\fwm & 0.01  & 9.60E-01 &\bfseries -0.50  & 4.10E-02 & -0.08 & 6.40E-01 &\bfseries -0.60  & \bfseries1.70E-04 & 0.19  & 2.90E-01 & -0.39 & 2.10E-02 \\
\sm & 0.04  & 8.70E-01 & -0.31 & 2.20E-01 & -0.12 & 4.90E-01 & -0.40  & 2.00E-02 & 0.16  & 3.80E-01 & -0.18 & 3.00E-01 \\
\dhb(mean) & 0.07  & 7.80E-01 &\bfseries -0.53 & 2.70E-02 & 0.07  & 7.00E-01 & \bfseries-0.57 & \bfseries3.90E-04 & 0.28  & 1.10E-01 & -0.42 & 1.40E-02 \\
\fwr &0.12  & 6.60E-01 & -0.36 & 1.60E-01 & -0.12 & 5.20E-01 & \bfseries-0.55 &\bfseries 8.30E-04 & 0.12  & 3.40E-01 & -0.36 & 5.30E-02 \\
\sr& 0.13  & 6.20E-01 & -0.29 & 2.60E-01 & -0.23 & 1.90E-01 & -0.47 & \bfseries5.70E-03 & 0.13  & 7.20E-01 & -0.29 & 1.80E-01 \\
\dhb(rms) & 0.05  & 8.50E-01 & -0.32 & 2.20E-01 & -0.02 & 9.10E-01 & -0.47 & \bfseries6.30E-03 & 0.05  & 7.30E-01 & -0.32 & 1.30E-02 \\
\rfe & 0.04  & 8.70E-01 & -0.14 & 6.20E-01 & 0.32  & 7.40E-02 & 0.32  & 7.20E-02 & -0.07 & 7.10E-01 & 0.02  & 9.00E-01 \\
$\tau$ & -0.18 & 4.90E-01 & -0.12 & 6.50E-01 & -0.25 & 1.60E-01 & -0.11 & 5.30E-01 & -0.13 & 4.70E-01 & -0.02 & 9.00E-01 \\
$\lv$ & -0.05 & 8.50E-01 & -0.01 & 9.70E-01 & -0.10  & 5.80E-01 & -0.03 & 8.80E-01 & 0.06  & 7.30E-01 & 0.12  & 5.10E-01 \\
\lb/\ledd &\bfseries-0.50  & 4.00E-02 & -0.13 & 6.30E-01 & -0.39 & 3.20E-02 & -0.13 & 4.60E-01 & \bfseries-0.63 & \bfseries6.70E-05 & -0.17 & 3.50E-01 \\
\hline
\multicolumn{13}{c}{RMS SPECTRUM} \\
\hline
 & \multicolumn{2}{c}{$f_{\rm \sigma, rms}$} & \multicolumn{2}{c}{$f_{\rm F, rms}$} & \multicolumn{2}{c}{$f_{\rm \sigma, rms}$} & \multicolumn{2}{c}{$f_{\rm F, rms}$}& \multicolumn{2}{c}{$f_{\rm \sigma, rms}$} & \multicolumn{2}{c}{$f_{\rm F, rms}$}  \\
 \hline
\sst& 0.11  & 6.80E-01 & 0.10   & 6.90E-01 & 0.09  & 6.20E-01 & 0.05  & 7.60E-01 & 0.11  & 1.40E-01 & 0.10   & 3.50E-01 \\
\fwm & -0.20  & 4.50E-01 & -0.46 & 6.40E-02 & -0.26 & 1.40E-01 & \bfseries-0.55 &\bfseries 8.30E-04 & -0.20  & 9.50E-01 & -0.46 & 3.50E-02 \\
\sm & -0.14 & 5.90E-01 & -0.28 & 2.80E-01 & -0.27 & 1.30E-01 & -0.43 & 1.20E-02 & -0.14 & 9.50E-01 & -0.28 & 2.50E-01 \\
\dhb(mean) & -0.14 & 6.00E-01 &\bfseries -0.51 & 3.70E-02 & -0.12 & 4.90E-01 & -0.47 & \bfseries6.40E-03 & -0.14 & 7.20E-01 & \bfseries-0.51 & 4.80E-02 \\
\fwr  & -0.12 & 6.60E-01 & -0.38 & 1.30E-01 & -0.35 & 4.90E-02 & \bfseries-0.63 &\bfseries 7.70E-05 & -0.12 & 8.40E-01 & -0.38 & 1.90E-02 \\
\sr & -0.18 & 4.90E-01 & -0.31 & 2.30E-01 & \bfseries-0.53 & \bfseries1.40E-03 & \bfseries-0.6  &\bfseries 2.50E-04 & -0.18 & 2.10E-01 & -0.31 & 5.10E-02 \\
\dhb(rms)& 0.24  & 3.50E-01 & -0.28 & 2.80E-01 & 0.05  & 7.80E-01 & -0.42 & 1.50E-02 & 0.24  & 4.20E-01 & -0.28 & 1.40E-02 \\
\rfe& 0.02  & 9.50E-01 & 0.02  & 9.40E-01 & 0.35  & 4.70E-02 & 0.39  & 2.80E-02 & 0.03  & 8.90E-01 & 0.10   & 5.50E-01 \\
$\tau$& -0.35 & 1.70E-01 & -0.15 & 5.70E-01 & -0.32 & 6.80E-02 & -0.20  & 2.60E-01 & -0.35 & 4.90E-01 & -0.15 & 6.60E-01 \\
$\lv$ &-0.22 & 4.10E-01 & -0.06 & 8.20E-01 & -0.20  & 2.60E-01 & -0.09 & 6.10E-01 & -0.22 & 8.80E-01 & -0.06 & 7.10E-01 \\
\lb/\ledd & -0.49 & 4.60E-02 & -0.20  & 4.30E-01 & -0.44 & \bfseries1.10E-03 & -0.17 & 3.30E-01 & -0.47 & \bfseries6.30E-03 & -0.20  & 2.60E-01 \\
\hline
\end{tabular}
\label{tab5}
\end{center}
\end{table*}

\begin{table*}
\begin{center}
\caption{The factor $f$ derived from the BLRs dynamical model, X-ray variability, resolved Pa$\alpha$ emission region  for 19 AGNs. }
\begin{tabular}{llllllll}
\hline
Name & $\tau$ & $\rm FWHM_{mean}$ & $\log \vpfm$ & $\log {\mbh}$ & $\log f_{\rm F,mean}$ & ref. \\
& (light days) & ($\kms$) & ($\msun$) &  ($\msun$) & & \\
\hline
Mrk     1310&$ 4.20^{+ 0.90}_{- 0.10}$&$ 2409\pm 24$&$ 6.68^{+ 0.09}_{- 0.01}$&$ 7.42^{+ 0.26}_{- 0.27}$&$ 0.74^{+ 0.26}_{- 0.28}$&P14,1\\
NGC     5548&$ 5.50^{+ 0.60}_{- 0.70}$&$12771\pm 71$&$ 8.24^{+ 0.05}_{- 0.05}$&$ 7.51^{+ 0.23}_{- 0.14}$&$-0.73^{+ 0.24}_{- 0.15}$&P14,1\\
NGC     6814&$ 7.40^{+ 0.10}_{- 0.10}$&$ 3323\pm  7$&$ 7.20^{+ 0.01}_{- 0.01}$&$ 6.42^{+ 0.24}_{- 0.18}$&$-0.78^{+ 0.24}_{- 0.18}$&P14,1\\
SBS 1116+583&$ 2.40^{+ 0.90}_{- 0.90}$&$ 3668\pm186$&$ 6.80^{+ 0.17}_{- 0.17}$&$ 6.99^{+ 0.32}_{- 0.25}$&$ 0.19^{+ 0.36}_{- 0.30}$&P14,1\\
Mrk      335&$14.10^{+ 0.40}_{- 0.40}$&$ 1273\pm 64$&$ 6.65^{+ 0.04}_{- 0.04}$&$ 7.25^{+ 0.10}_{- 0.10}$&$ 0.60^{+ 0.11}_{- 0.11}$&G17,2\\
Mrk     1501&$15.50^{+ 2.20}_{- 1.80}$&$ 3494\pm 35$&$ 7.57^{+ 0.06}_{- 0.05}$&$ 7.86^{+ 0.20}_{- 0.17}$&$ 0.29^{+ 0.21}_{- 0.18}$&G17,2\\
3C       120&$27.20^{+ 1.10}_{- 1.10}$&$ 1430\pm 16$&$ 7.04^{+ 0.02}_{- 0.02}$&$ 7.84^{+ 0.14}_{- 0.19}$&$ 0.80^{+ 0.14}_{- 0.19}$&G17,2\\
PG  2130+099&$35.00^{+ 5.00}_{- 5.00}$&$ 1781\pm  5$&$ 7.34^{+ 0.06}_{- 0.06}$&$ 6.92^{+ 0.24}_{- 0.23}$&$-0.42^{+ 0.25}_{- 0.24}$&G17,1\\
Arp      151&$ 3.60^{+ 0.70}_{- 0.20}$&$ 3098\pm 69$&$ 6.83^{+ 0.09}_{- 0.03}$&$ 6.66^{+ 0.26}_{- 0.17}$&$-0.17^{+ 0.26}_{- 0.19}$&P18,1\\
Arp      151&$ 5.61^{+ 0.66}_{- 0.84}$&$ 2021\pm 17$&$ 6.65^{+ 0.05}_{- 0.06}$&$ 6.93^{+ 0.33}_{- 0.16}$&$ 0.28^{+ 0.34}_{- 0.17}$&P18,1\\
Arp      151&$ 7.52^{+ 1.43}_{- 1.06}$&$ 2872\pm 90$&$ 7.08^{+ 0.09}_{- 0.07}$&$ 6.92^{+ 0.50}_{- 0.23}$&$-0.16^{+ 0.50}_{- 0.25}$&P18,3\\
Mrk       50&$ 8.66^{+ 1.63}_{- 1.51}$&$ 4101\pm 56$&$ 7.45^{+ 0.08}_{- 0.08}$&$ 7.50^{+ 0.25}_{- 0.18}$&$ 0.05^{+ 0.26}_{- 0.20}$&W18,4 \\
PG  1310-108&$ 4.50^{+ 1.40}_{- 1.00}$&$ 3422\pm 21$&$ 7.01^{+ 0.13}_{- 0.10}$&$ 6.48^{+ 0.21}_{- 0.18}$&$-0.53^{+ 0.23}_{- 0.22}$&W18,4\\
Mrk      141&$ 5.63^{+ 8.27}_{- 1.65}$&$ 5129\pm 45$&$ 7.46^{+ 0.63}_{- 0.13}$&$ 7.46^{+ 0.15}_{- 0.21}$&$ 0.00^{+ 0.20}_{- 0.66}$&W18,4\\
PG  1351+695&$16.00^{+ 5.30}_{- 5.60}$&$ 4099\pm 43$&$ 7.72^{+ 0.14}_{- 0.15}$&$ 7.58^{+ 0.08}_{- 0.08}$&$-0.14^{+ 0.17}_{- 0.16}$&W18,4\\
Mrk     1511&$ 5.44^{+ 0.74}_{- 0.67}$&$ 4154\pm 28$&$ 7.26^{+ 0.06}_{- 0.05}$&$ 7.11^{+ 0.20}_{- 0.17}$&$-0.15^{+ 0.21}_{- 0.18}$&W18,4\\
NGC     4593&$ 3.54^{+ 0.76}_{- 0.82}$&$ 4264\pm 41$&$ 7.10^{+ 0.09}_{- 0.10}$&$ 6.65^{+ 0.27}_{- 0.15}$&$-0.45^{+ 0.29}_{- 0.17}$&W18,4\\
ZW    229-15&$ 3.86^{+ 0.69}_{- 0.90}$&$ 3705\pm203$&$ 7.01^{+ 0.09}_{- 0.11}$&$ 6.94^{+ 0.14}_{- 0.14}$&$-0.07^{+ 0.18}_{- 0.17}$&W18,4\\
Mrk      142&$ 7.90^{+ 1.20}_{- 1.10}$&$ 1588\pm 58$&$ 6.59^{+ 0.07}_{- 0.07}$&$ 6.23^{+ 0.26}_{- 0.45}$&$-0.36^{+ 0.27}_{- 0.46}$&L18,5\\
1H  0323+342&$14.80^{+ 3.90}_{- 2.70}$&$ 1388\pm 32$&$ 6.75^{+ 0.12}_{- 0.08}$&$ 6.68^{+ 0.22}_{- 0.22}$&$-0.07^{+ 0.23}_{- 0.25}$&X18,6\\
3C        273&$ 146.8^{+  8.3}_{- 12.1}$&$ 3314\pm 59$&$ 8.50^{+ 0.03}_{- 0.04}$&$ 8.41^{+ 0.23}_{- 0.23}$&$-0.09^{+ 0.23}_{- 0.23}$&S18,7\\
\hline
\end{tabular}
\label{tab6}
\end{center}
References: The $\log\mbh$ measured by BLR dynamical model, X-ray variability and resolve the Pa$\alpha$ emission region from P14, G17, P18, W18, L18, X18, S18 which come from \cite{Pancoast2014,Gr17a,Williams2018,Pancoast2018,Li2018,Pan2018,Sturm2018}, respectively. The references of $\tau$ and \fwm  come from (1) Table \ref{tab1} in this paper; (2) \cite{Gr12};  (3) \cite{Va15};  (4) \cite{Ba15}; (5) \cite{Hu15}; (6) \cite{Wang2016}; (7) \cite{Zhang2019}

\end{table*}

\end{document}